%% file: main.tex
\documentclass[sigconf]{acmart}

\usepackage{geometry}
\usepackage{booktabs} 
\usepackage{subcaption}
\usepackage{courier} 
\usepackage{alphalph}

\usepackage{subcaption}

\begin{document}
\title{Memory Centric Characterization and Analysis of SPEC CPU2017 Suite\textsuperscript{1}}

\author{Sarabjeet Singh}
\email{sarabjeet.singh@ashoka.edu.in}
\orcid{0000-0003-3032-1916}
\affiliation{%
  \institution{Ashoka University}
}
\author{Manu Awasthi}
\email{manu.awasthi@ashoka.edu.in}
\affiliation{%
  \institution{Ashoka University}
}

\renewcommand{\shortauthors}{Singh and Awasthi}

\begin{abstract}
\input{abstract}
\end{abstract}


\maketitle
\footnotetext[1]{A condensed version of this work was published in ICPE 2019 \cite{singh2019memory} (\href{https://doi.org/10.1145/3297663.3310311}{https://doi.org/10.1145/3297663.3310311}).}

\input{intro}

\input{background}
\input{methodology}
\input{InstructionAnalysis}
\input{MemoryAnalysis}
\input{NewBenchmarks}
\input{related}
\input{conclusion}

\begin{acks}
The authors thank the anonymous reviewers and shepherd for their useful comments and feedback. This work was supported in part by SERB grant (ECR/2017/000887), as well as Indian Institute of Technology Gandhinagar and Ashoka University, Sonipat.
\end{acks}

\bibliographystyle{ACM-Reference-Format}
\bibliography{bibliography}

\end{document}

%% file: abstract.tex
In this paper we provide a comprehensive, memory-centric characterization
 of the SPEC CPU2017 benchmark suite, using a number of mechanisms including
 dynamic binary instrumentations, measurements on native hardware using
 hardware performance counters and OS based tools.

We present a number of results including working set sizes, memory capacity
consumptions and, memory bandwidth utilization of various workloads.
Our experiments reveal that the SPEC CPU2017 workloads are surprisingly memory 
intensive, with
approximately 50\% of all dynamic instructions being memory intensive 
ones. We also show that there is a large variation in the memory 
footprint and bandwidth utilization profiles of the entire suite, with 
some benchmarks using as much as 16 GB of main memory and up to 
2.3 GB/s of memory bandwidth. 

We also perform instruction execution and distribution analysis of the 
 suite and find that the average instruction count for SPEC CPU2017 workloads
is an order of magnitude higher than SPEC CPU2006 ones. In addition, we also 
find that FP benchmarks of the SPEC 2017 suite have higher compute 
requirements: on  average, FP workloads execute
three times the number of compute operations as compared to INT workloads.

%% file: intro.tex
\section{Introduction}\label{sec:introduction}

\scriptsize
\begin{table*}[hbtp]
\centering
\caption{SPEC CPU2017 Benchmarks}
\label{table:BenchmarkDescription}
\begin{tabular}{|c|c|c|}
\hline
\textbf{Benchmark} & \textbf{Domain}                & \textbf{Inputs}                                                                     \\ \hline
perlbench          & Perl Interpreter               & Interprets SpamAssassin, MHonArc, and an internal script                            \\ \hline
gcc                & GNU C Compiler                 & C source code file compiled with different optimizations                            \\ \hline
mcf                & Route Planning                 & Solves large-scale minimum cost flow problem                                        \\ \hline
omnetpp            & Discrete Event Simulation      & Simulates a 10 Gb Ethernet network                                                  \\ \hline
xalancbmk          & XML to HTML Conversion         & XML documents to HTML, text, and other XML types                                    \\ \hline
x264               & Video Compression              & Compresses portions of Blender Open Movie Project's "Big Buck Bunny"                \\ \hline
deepsjeng          & Artificial Intelligence        & Plays Chess variants employing alpha-beta tree search                               \\ \hline
leela              & Artificial Intelligence        & Computer Go featuring Monte Carlo tree search                                       \\ \hline
exchange2          & Artificial Intelligence        & Recursively solves 27 9x9 Sudoku puzzles                                            \\ \hline
xz                 & General Data Compression       & Tar archive, database of ClamAV signatures and a file with text and image data      \\ \hline
bwaves             & Explosion Modeling             & Simulates blast waves in 3D                                                         \\ \hline
cactuBSSN          & Physics: Relativity            & Uses BSSN formulation of Einstein equation and employs finite differencing in space \\ \hline
namd               & Molecular Dynamics             & 92224 atom simulation of Apolipoprotein A-I                                         \\ \hline
parest             & Biomedical Imaging             & 3D reconstruction of interior of a body using multiple 2D observations              \\ \hline
povray             & Ray Tracing                    & Renders a 2560x2048 pixel image of a chess board with pieces                        \\ \hline
lbm                & Fluid Dynamics                 & Simulates flow of an incompressible fluid in 3D                                     \\ \hline
wrf                & Weather Forecasting            & Simulates the January 2000 North American Blizzard                                  \\ \hline
blender            & 3D Rendering and Animation     & Simulates reduced version of the Weybec Crazy Glue shot 3 data set to image         \\ \hline
cam4               & Atmosphere Modeling            & Atmospheric Component of the NCAR Earth System                                      \\ \hline
pop2               & Wide-scale Ocean Modeling      & Ocean Component of the NCAR Earth System                                            \\ \hline
imagick            & Image Manipulation             & Performs various transformations on input images                                    \\ \hline
nab                & Molecular Dynamics             & Models molecules with varying number of atoms                                       \\ \hline
fotonik3d          & Computational Electromagnetics & Employs finite-difference time-domain method for the Maxwell equations              \\ \hline
roms               & Regional Ocean Modeling        & A free-surface, hydrostatic, primitive equation model                               \\ \hline
\end{tabular}
\end{table*}
\normalsize


\par The study of computer architecture and system design depends on the availability of new workloads that are able to faithfully represent the contemporary and future applications of a given vertical. In the CPU design domain Standard Performance Evaluation Corporation (SPEC) has been releasing the SPEC CPU suite for close to three decades now, starting in 1992. These benchmarks have become the standard for any researcher or commercial entity wishing to benchmark their architecture or design against existing ones.

\par The latest offering of SPEC CPU suite, SPEC CPU 2017, was released in June 2018~\cite{CPU2017_Description}, ending a decade long wait for a new set of CPU benchmarks -- the version before this was released in 2006. SPEC CPU 2017 retains a number of the benchmarks from previous iterations, but has also added many new ones to reflect the changing nature of applications. The 2017 suite has added a lot of excitement in the community with researchers already working on characterizing the set of workloads to figure out system bottlenecks~\cite{CPU2017_characterization,CPU2017_WaitofaDecade}.

\par In the last decade, main memory, typically made of Dynamic Random Access Memory (DRAM) in most contemporary systems, has become a first class component of the compute system design space. The last decade has seen a renewed interest in the architectural design space exploration of main memory, including novel additions to the existing interfaces and architecture (JEDEC, DDR3, DDR4, DDR5)~\cite{ahn2015pim,Udipi_DRAM,liu2012raidr,hassan2016chargecache}. Not only this, exploration of emerging memory technologies like Phase Change Memory, MRAM etc., to find their space in the memory hierarchy has also been carried out~\cite{lee2009architecting,jin2014area}. To that end, a number of works have carried out these explorations for using emerging memories in both the cache and main memory architectures~\cite{awasthi2012managing,zhang2009exploring,kultursay2013evaluating,xu2011design,wu2009hybrid}.

\par SPEC 2006~\cite{CPU2006} has played a very important part in these explorations. The SPEC suites have had a set of memory intensive workloads (e.g. {\tt mcf}). Innovations to cache and memory hierarchies have been tested using these workloads by either (i) selecting individual workloads from the suite, or (ii) creating workload mixes, with varying types of memory behavior (also known as multi-programmed workloads). This was made possible by the already characterized memory behavior patterns of different workloads in the SPEC 2006 suite~\cite{CPU2006_Jaleel}.

\par However, with SPEC CPU2017 being a relatively new offering, there is no existing work that characterizes the memory hierarchy behavior of the suite. In this work, we wish to bridge this existing gap in literature. As a result, we make the following important contributions in this paper:

\begin{enumerate}
	\item Across the SPEC CPU2017 suite, we provide a holistic characterization 
	of the dynamic instruction execution profiles of different workloads, for 
	both Rate and Speed categories, and observe that most workloads have a 
	large number of memory related operations: close to 50\% across the suite.
	We also provide an opcode level classification of workloads.
	\item Most importantly, we provide a detailed analysis of the memory 
	behavior of various benchmarks, using a combination of dynamic 
	instrumentation tools (Pin/Pintools), hardware performance counters and
	operating system level tools to report the overall working set size, memory 
	bandwidth consumption and memory resident working set sizes of different 
	workloads, respectively.
\end{enumerate}

The rest of the paper is organized as follows. Section~\ref{section:SPEC} 
gives a background of CPU2017 benchmarks. Section~\ref{section:methodology} 
proposes the methodology used to characterize the benchmarks. 
Section~\ref{section:instructionprofile} and \ref{section:MemoryBehavior} 
analyses the benchmarks at an instruction and memory level, respectively. 
Section~\ref{section:NewBenchmarks} discusses the benchmarks which are newly 
added to the SPEC CPU suite. Finally, we discuss the related works in 
Section~\ref{section:related_work} and conclude in 
Section~\ref{section:conclusion}.

%% file: background.tex
\section{SPEC CPU2017}\label{section:SPEC}

SPEC CPU is a widely acknowledged suite of compute intensive benchmarks, which tests processor's, memory system's and compiler's performance. A number of versions of SPEC have been released over the years, with the latest version, released in 2017, and aptly named, SPEC CPU2017. CPU2017~\cite{CPU2017} considers state-of-the-art applications, organizing 43 benchmarks into four different sub-suites: 10 rate integer (INTRate), 10 speed integer (INTSpeed), 13 rate floating point (FPRate) and 10 speed floating point (FPSpeed). The speed and rate suites vary in workload sizes, compile flags and run rules. SPEC\textit{speed} measures the time for completion by running a single copy of each benchmark, with an option of using multiple OpenMP threads. Hence, speed is a measure of single thread performance, typically measured by metrics like IPC (Instructions Per Cycle). On the other hand, SPEC\textit{rate} measures the throughput of the overall chip, with possibly multiple cores, by running multiple, concurrent copies of the \textit{same} benchmark with OpenMP disabled. Most applications have both rate and speed versions (denoted as \textit{5nn.benchmark\_r} and \textit{6nn.benchmark\_s}, respectively), except for \textit{namd}, \textit{parest}, \textit{povray} and \textit{blender}, which only have the rate versions, and \textit{pop2}, which only has the speed version. Similar to SPEC CPU2006, SPEC CPU2017 has been provided with three input sets: \textbf{test} (to test if executables are functional), \textbf{train} (input set built using feedback-directed optimization and used for training binaries), and \textbf{ref} (timed data set of the real applications, which is intended for a reportable run). 
\par CPU2017 benchmarks and their input sets are described in Table~\ref{table:BenchmarkDescription}. Benchmarks from \textit{perlbench} to \textit{xz} are integer, while the rest are floating point applications. Workloads are modified to compress I/O operations other than file read/write as the intent is to measure the compute intensive portion of a real application, while minimizing I/O, thereby focusing on the performance of the CPU, memory and the compiler. SPEC CPU benchmarks are distributed as source code and must be compiled, which raises the question that how must they be compiled? There exists many possibilities ranging from \textit{debug, no-optimize} to a highly customized optimization. Any point in between would result in different results than others. For CPU2017, SPEC has chosen to allow two points in the range: \textbf{base}, where all modules of a given language in a suite must be compiled using the same flags, in the same order and \textbf{peak} which allows greater flexibility in order to achieve better performance. For peak, different compiler options may be used for each benchmark, and feedback-directed optimization is allowed.

%% file: methodology.tex
\section{Methodology}
\label{section:methodology}

\scriptsize
\begin{table}
\centering
\caption{System Configuration}
\label{table:CAD_config}
\begin{tabular}{|c|c|}
\hline
Model & 40-core Intel Xeon E5-2698 v4\\ \hline
CPU Frequency &  2.2GHz    \\ \hline
L1i cache & 8-way, 32 KB \\ \hline
L1d cache & 8-way, 32 KB \\ \hline
L2 cache  & 8-way, 256 KB \\ \hline
L3 cache & Shared 20-way, 50 MB \\ \hline
Cache line size & 64 Bytes \\ \hline
Main Memory & 505 GB, DDR4\\ \hline
Dynamic Frequency Scaling & On\\ \hline
\end{tabular}
\end{table}
\normalsize

\begin{figure*}
  \centering
  \hspace*{-10 pt}
  \includegraphics[width=2.15\columnwidth]{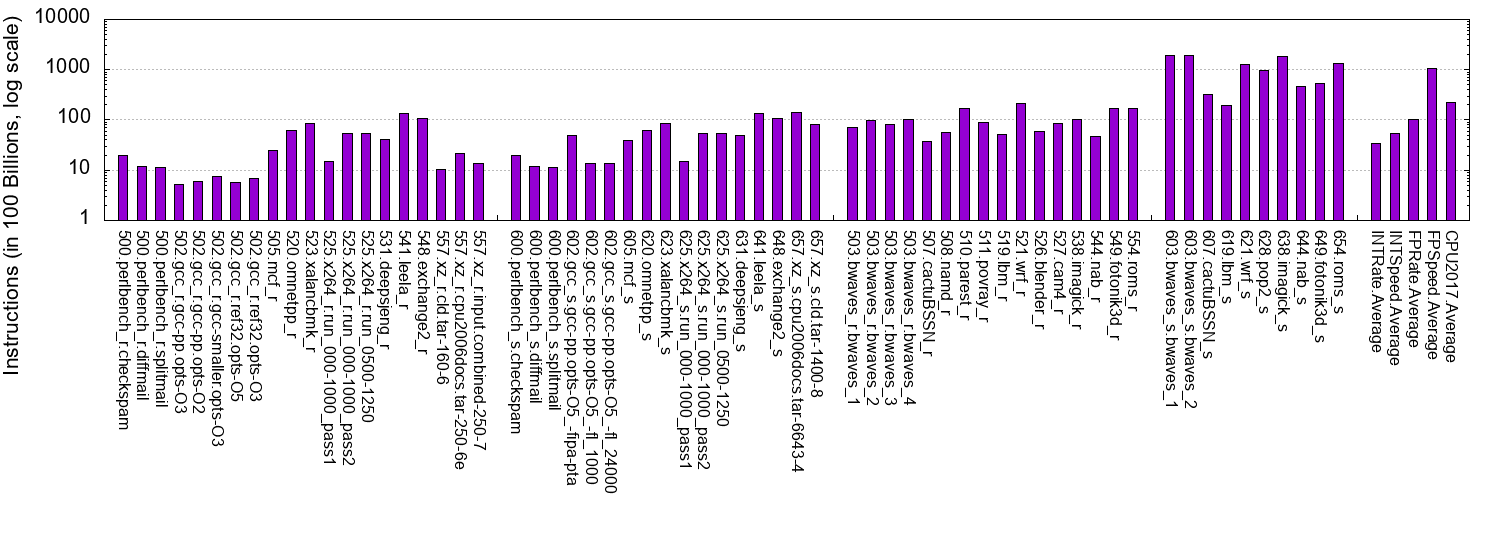}
  \vspace*{-40 pt}
  \caption{Dynamic Instruction Count. (Clusters from L to R: INTRate, INTSpeed, FPRate, FPSpeed)}
  \label{fig:DynamicInstructionCount}
\end{figure*}

To study the characteristics of CPU2017 workloads, we used a number of tools to analyse their behavior. The analysis in this paper is based on the x86\_64 instruction set. The binaries for the workloads were created using the default, SPEC recommended compiler flags~\cite{CPU2017_Compilation} ({\tt gcc -O3}), using compilation scripts which ship with CPU 2017.
Speed workloads are compiled to use 4 OpenMP threads, while rate workloads were executed with a single instance of the benchmark. We use Pin~\cite{Pin}, a dynamic binary instrumentation framework available for both 32 and 64 bit versions of the instruction set. Pin provides a rich set of APIs that can be used to study various characteristics of program behavior at the Instruction Set Architecture (ISA) level. These APIs are used to create a number of tools, called Pintools, with each capable of carrying out a certain type of analysis. In this study, we use the following pintools: \textit{ldst} (dynamic register/memory operand pattern profiler), \textit{opcodemix} (dynamic opcode mix profiler), and \textit{dcache} (a functional simulator of data cache).

For gathering information about workload behavior with real hardware, we use \textit{perf}, a performance analysis tool~\cite{Perf}, and \textit{ps}~\cite{ps}, an OS utility to collect process level memory related data for various workloads. Table~\ref{table:CAD_config} presents the configuration of the machine used to run experiments for Pin-based, hardware-counter and system-level experimentation and data collection. All the benchmarks were executed till completion. \textit{627.cam4\_s} faced a runtime error and hence is deprecated from this study.

%% file: InstructionAnalysis.tex
\section{Instruction Profile}
\label{section:instructionprofile} 

\begin{figure*}
  \centering
  \begin{subfigure}{2.15\columnwidth}
    \includegraphics[width=\textwidth]{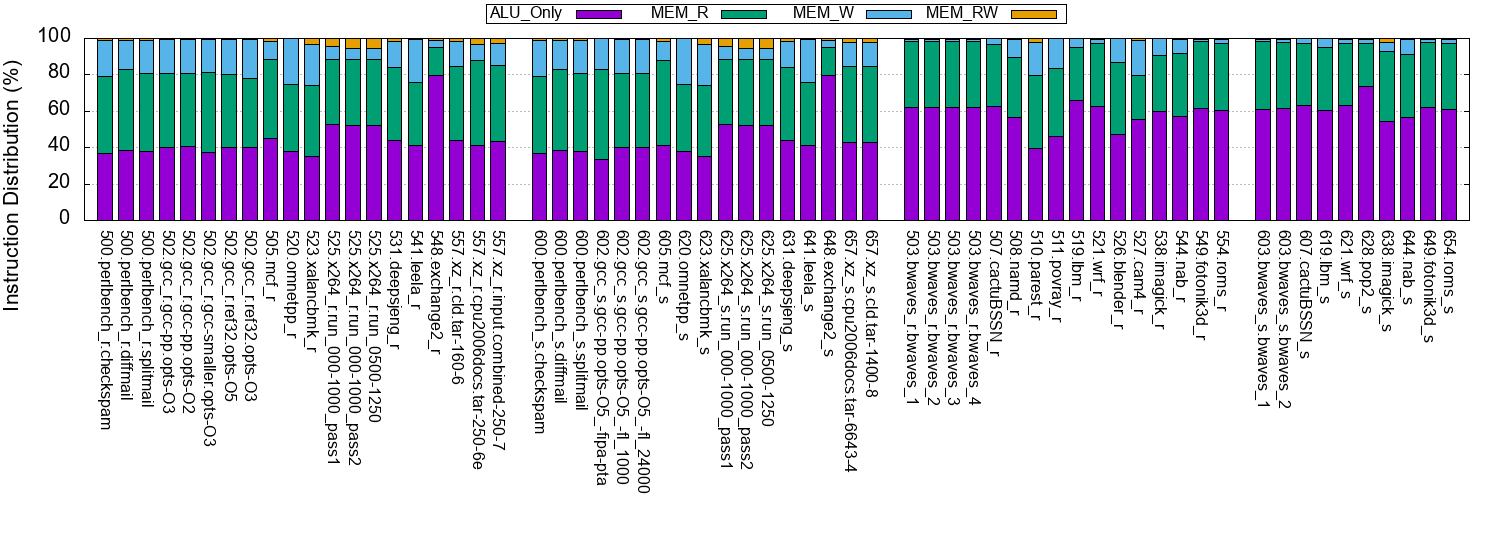}
  \vspace*{-35 pt}
    \caption{Instruction Distribution}
    \label{fig:InstructionMix}
  \end{subfigure}
  \begin{subfigure}{2.10\columnwidth}
    \includegraphics[width=\textwidth]{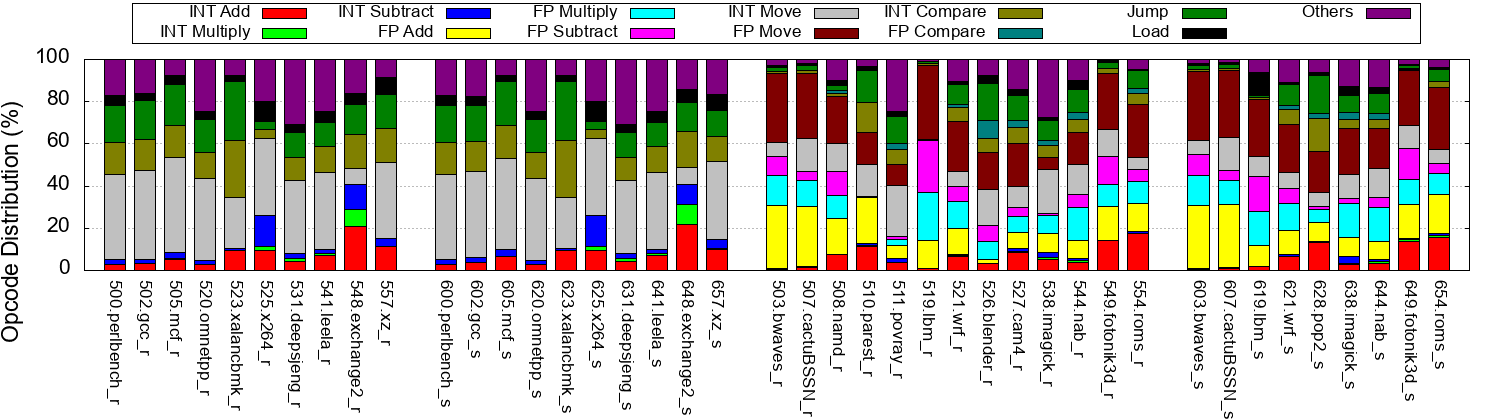}
  \vspace*{-15 pt}
    \caption{Opcode Distribution}
    \label{fig:OpcodeMix}
  \end{subfigure}
    \vspace*{-10 pt}
  \caption{Instruction Profile. (Clusters from L to R: INTRate, INTSpeed, FPRate, FPSpeed)}
  \label{fig:InstructionProfile}
\end{figure*}

\begin{figure*}[h]
  \centering
  \hspace*{-10 pt}
  \includegraphics[width=2.15\columnwidth]{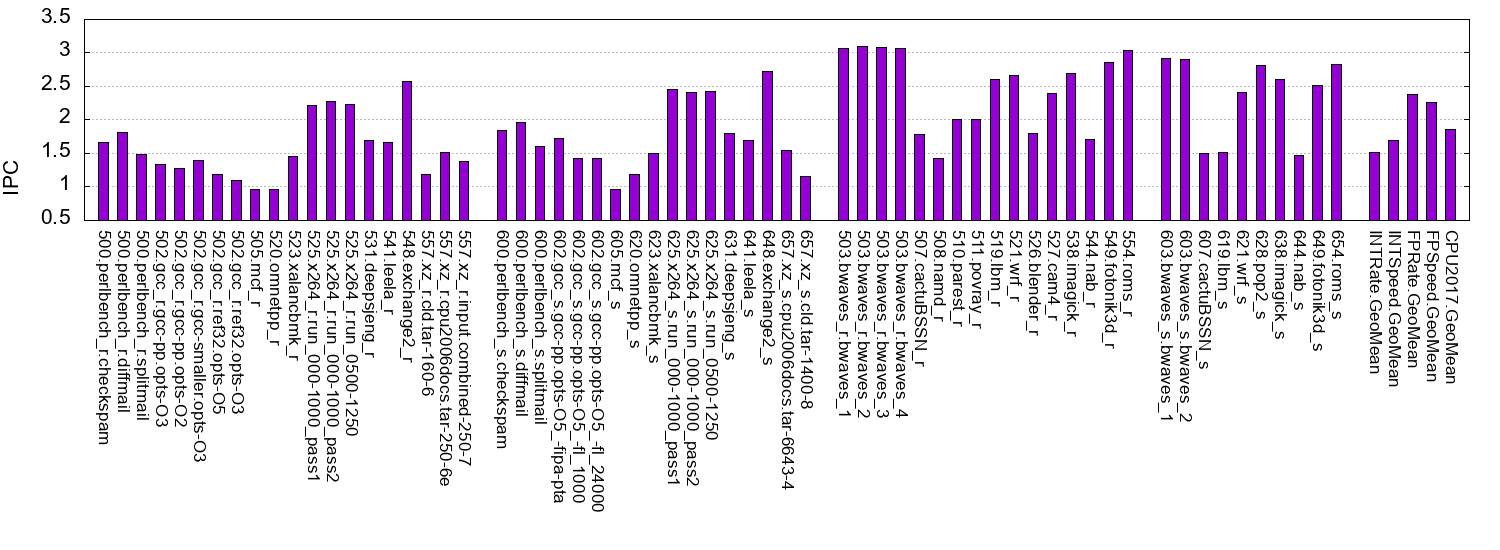}
  \vspace*{-35 pt}
  \caption{IPC. (Clusters from L to R: INTRate, INTSpeed, FPRate, FPSpeed)}
  \label{fig:IPC}
\end{figure*}

Analysis of the instruction profiles is a good mechanism for understanding a benchmark's behavior and locating potential sources of bottlenecks in hardware and software. To that end, we first study the dynamic instruction count, instruction distribution and the runtime performance of CPU 2017 workloads. 
\paragraph{Dynamic Instruction Count:} Figure~\ref{fig:DynamicInstructionCount} depicts the dynamic instruction count for each benchmark in the suite. Each benchmark is divided into its various input workloads, and presented in order of their sub-suites, with an indication of sub-suite average at the end. These results have been obtained using perf's hardware event \textit{instructions}. We note that the average instruction count for the SPEC CPU2017 is 22.19 trillion (1.4 Quadrillion in total) which is about an order of magnitude higher than the SPEC CPU2006~\cite{CPU2006,CPU2006_Jaleel}. It is observed that the FPSpeed suite have massive dynamic instruction count with respect to others, with \textit{bwaves\_s} reaching as high as 382 trillion. In general, Speed workloads have 1.5-10 times more instructions than the corresponding Rate ones, and floating point (FP) workloads have 3-17 times than the integer (INT) workloads. 

\par \textit{Dynamic Instruction Count:} Figure~\ref{fig:DynamicInstructionCount} depicts the dynamic instruction count, a count of total number of instructions executed by the workload. Each benchmark is divided into its constituent workloads, depending on the input set. These results were collected using perf's hardware event \textit{instructions}. We note that the average instruction count for SPEC CPU2017 workloads is 22.19 trillion (1.4 quadrillion in total) which is an order of magnitude higher than the SPEC CPU2006~\cite{CPU2006,CPU2006_Jaleel}. We also observe that the FPSpeed suite has a much larger dynamic instruction count with respect to others sub-suites, with \textit{bwaves\_s} executing
as many as 382 trillion instructions. In general, Speed workloads have 1.5-10 $\times$ more instructions than the corresponding Rate ones, and floating point (FP) workloads have 3-17 times than the integer (INT) workloads. These observations point to the general increase in the complexity of SPEC CPU workloads over the years.
\par \textit{Instruction Distribution:} To better understand the distribution of instructions that access memory, we present the instruction distribution for workloads in Figure~\ref{fig:InstructionMix}. These experiments were conducted using the \textit{ldst} Pintool. Some benchmarks like \textit{perlbench}, \textit{x264}, \textit{bwaves} and a few others have multiple input files, which are executed in succession to complete the run. We report the results of each of these runs individually, leading to multiple bars for a benchmark. To keep the discussion simple, we divide the instructions into four broad categories: instructions that do not refer memory (called ALU Only in the figure), instructions that have one or more source operands in memory (called MEM\_R), instructions whose the destination operand is in memory (MEM\_W), and instructions whose source and destination operands are in memory (MEM\_RW)\footnote{Memory-to-memory instructions like \textit{movs} in x86 are billed under the MEM\_RW bucket.}. 

This broad classification allows us to compare the types of instructions that are executed by each benchmark, and provides a first order insight into the memory behavior of these benchmarks. We make a few interesting observations. First, irrespective of the input sets provided, the instruction distribution of a benchmark, across these four buckets doesn't change drastically. This is evidenced by the instruction distributions of all benchmarks that have multiple input files (\textit{perlbench}, \textit{gcc}, \textit{x264}, \textit{xz}, \textit{bwaves}). Also, the instruction distribution across the four buckets doesn't change significantly, irrespective of whether the speed or the rate version of the benchmark is being examined. 
 
Most benchmarks have a fairly balanced percentage of instructions that fall under either one of the MEM\_R/ MEM\_W/ MEM\_RW or the ALU\_Only 
buckets. However, a few exceptions like \textit{exchange2} (AI) and \textit{pop2} (Ocean Modeling) exist where the contribution of ALU\_Only operations is fairly significant at 79.6\% and 73.5\%, respectively. 
Floating point workloads also exhibit a lot of compute activity, with $\sim$60\% ALU\_Only instructions. However, on an average across the 
benchmark suite, SPECInt sub-suite exhibits executes more memory related instructions than the SPECFP one. Our observations are consistent with the earlier versions of SPEC: CPU2006 and CPU2000~\cite{CPU2006_Jaleel}. 

In order to get insights regarding the type of operations done by these instructions, we profile the benchmarks to report instruction level classification. Results, collected  with the help of \textit{opcodemix} pintool, are presented in Figure~\ref{fig:OpcodeMix}. The results for one benchmark were averaged across all their input files. 
We observe that FP workloads, have approximately three times the number of arithmetic operations than the INT workloads. In addition, we observe that a majority of the memory operations in both integer and floating point sub-suites are dominated by their respective \text{move} operations.
We also observe that memory instructions for both Int and FP benchmarks are predominantly read-only, which is consistent with the high-level results obtained in Figure~\ref{fig:InstructionMix}.

\textit{Performance:} We report the performance of the workloads in terms of instruction per cycle (IPC) in Figure~\ref{fig:IPC}, on the system outlined in Table~\ref{table:CAD_config}. IPC is calculated as the ratio of the hardware events \textit{instructions} and \textit{cpu-cycles}, obtained using \textit{perf}. To account for variations in execution time due to variables that cannot be controlled, each experiment is run three times and the average values are reported. Order of benchmark execution is shuffled between repetitions to mitigate measurement bias. Rest of the experiments in the paper are not repeated. We observe that FP workloads have better IPC than INT ones. 
However, we do note that applications that execute a significant number of memory related operations (e.g. \textit{cactuBSSN\_s}, \textit{lbm\_s}, \textit{xz\_.cld.tar-1400-8} and \textit{mcf}) and have larger working sets, requiring more accesses to the memory hierarchy, have lower IPCs.

%% file: MemoryAnalysis.tex
\section{Memory Behavior}
\label{section:MemoryBehavior}

\subsection{Spatial Locality Behavior}
\label{section:MemoryReferenceSize}
\begin{figure}
\centering
\includegraphics[width=0.45\textwidth,height=0.25\textwidth]{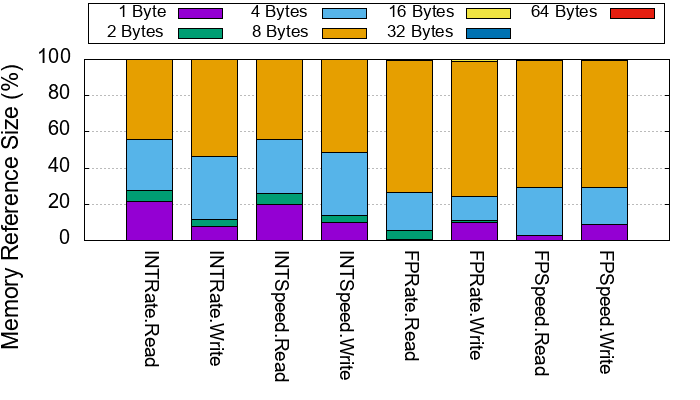}
  \vspace*{-15pt}
  \caption{Memory Reference Size}
  \label{fig:MemoryReferenceSize}
\end{figure}

Next, we observe the spatial locality characteristics of the workloads by observing benchmarks using \textit{opcodemix} Pintool. Opcodemix helps analyse  of the amount and size of data touched by any memory operation
 that requires to traverse the cache and memory hierarchy. We classify the instructions based on the amount of data that they access during these operations. 
In the interest of space, we present results averaged across the suites in Figure~\ref{fig:MemoryReferenceSize}. There is a broad range of data size granularities accessed by instructions, from 1~Byte to 64~Bytes, with the latter being the cacheline size as well. However, two important figures stand out. First, the majority of the accesses (64\%) are for an exact 8~Byte granularity. Second, 99.5\% of accesses (reads and writes) are for 8~Bytes or smaller access granularities. The number of accesses to larger data granularities is extremely small, and holds true across the suite. This indicates limited spatial data locality at the individual instruction level.


\input{WorkingSet.tex}

\input{MemoryConsumption.tex}
\input{MemoryBandwidth.tex}

%% file: WorkingSet.tex
\subsection{Working Set Sizes}
\label{section:WorkingSet}

\captionsetup[sub]{font=scriptsize}
\begin{figure*}[hbtp]
\vspace*{-30 pt}
\centering
\begin{subfigure}{0.5\columnwidth}
	\includegraphics[width=\textwidth]{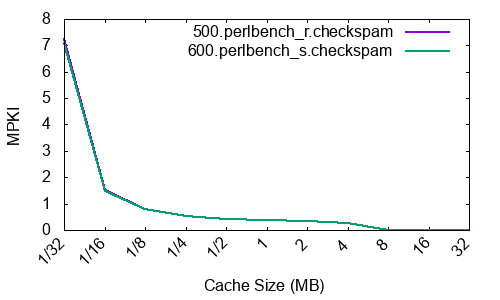}
  \caption{perlbench}
\end{subfigure}\hfill
\begin{subfigure}{0.5\columnwidth}
	\includegraphics[width=\textwidth]{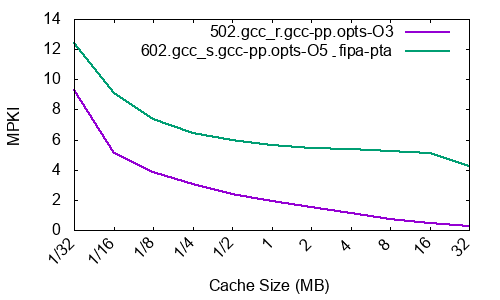}
  \caption{gcc}
\end{subfigure}\hfill
\begin{subfigure}{0.5\columnwidth}
	\includegraphics[width=\textwidth]{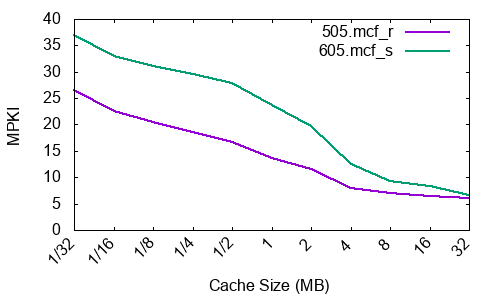}
  \caption{mcf}
\end{subfigure}\hfill
\begin{subfigure}{0.5\columnwidth}
	\includegraphics[width=\textwidth]{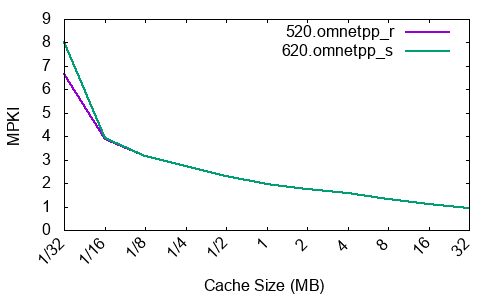}
  \caption{omnetpp}
\end{subfigure}\hfill
\begin{subfigure}{0.5\columnwidth}
	\includegraphics[width=\textwidth]{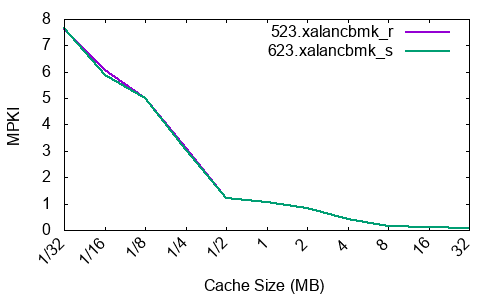}
  \caption{xalancbmk}
\end{subfigure}\hfill
\begin{subfigure}{0.5\columnwidth}
	\includegraphics[width=\textwidth]{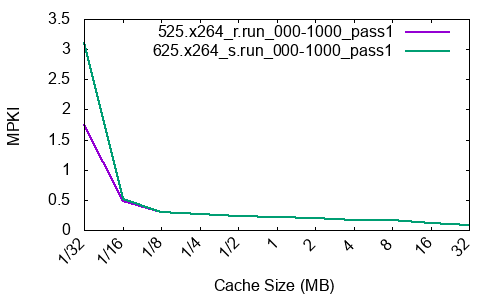}
  \caption{x264}
\end{subfigure}\hfill
\begin{subfigure}{0.5\columnwidth}
	\includegraphics[width=\textwidth]{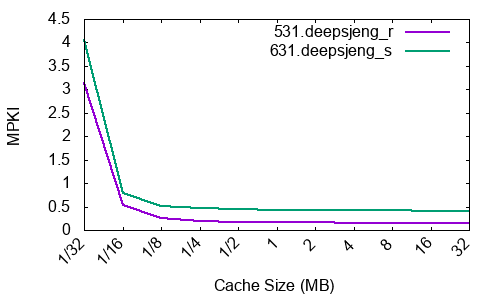}
  \caption{deepsjeng}
\end{subfigure}\hfill
\begin{subfigure}{0.5\columnwidth}
	\includegraphics[width=\textwidth]{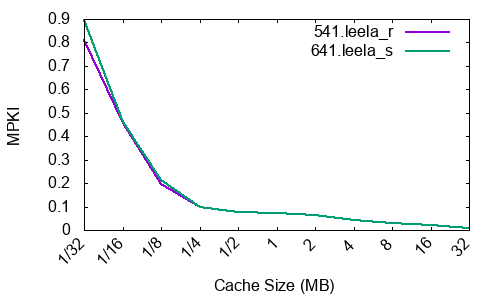}
  \caption{leela}
\end{subfigure}\hfill
\begin{subfigure}{0.5\columnwidth}
	\includegraphics[width=\textwidth]{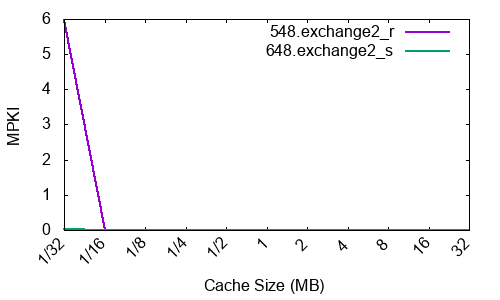}
  \caption{exchange2}
\end{subfigure}\hfill
\begin{subfigure}{0.5\columnwidth}
	\includegraphics[width=\textwidth]{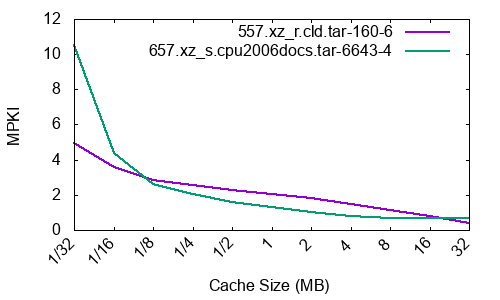}
  \caption{xz}
\end{subfigure}\hfill
\begin{subfigure}{0.5\columnwidth}
	\includegraphics[width=\textwidth]{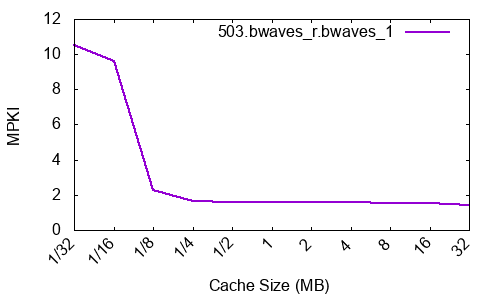}
  \caption{bwaves}
\end{subfigure}\hfill
\begin{subfigure}{0.5\columnwidth}
	\includegraphics[width=\textwidth]{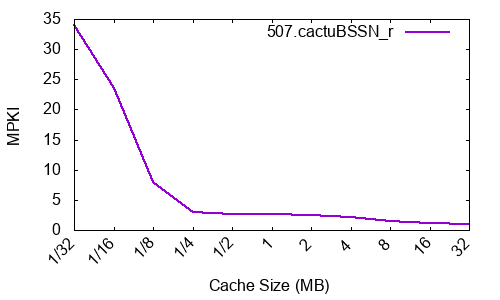}
  \caption{cactuBSSN}
\end{subfigure}\hfill
\begin{subfigure}{0.5\columnwidth}
	\includegraphics[width=\textwidth]{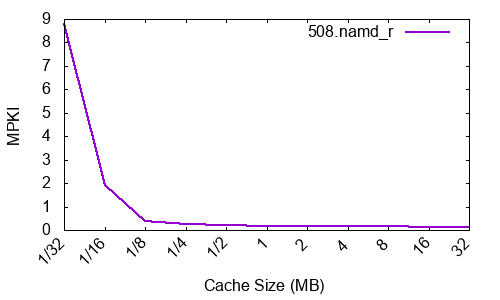}
  \caption{namd}
\end{subfigure}\hfill
\begin{subfigure}{0.5\columnwidth}
	\includegraphics[width=\textwidth]{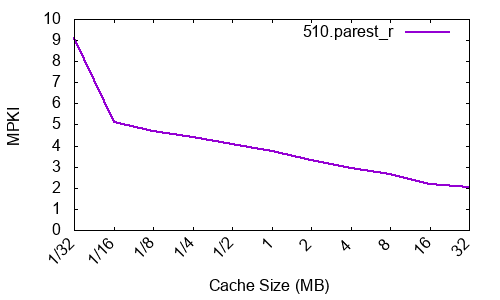}
  \caption{parest}
\end{subfigure}\hfill
\begin{subfigure}{0.5\columnwidth}
	\includegraphics[width=\textwidth]{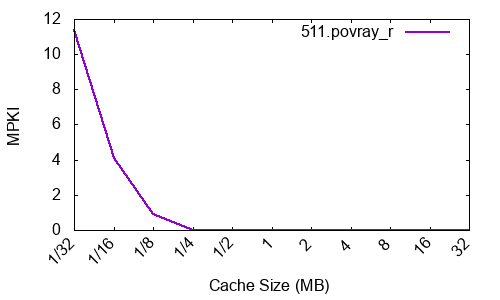}
  \caption{povray}
\end{subfigure}\hfill
\begin{subfigure}{0.5\columnwidth}
	\includegraphics[width=\textwidth]{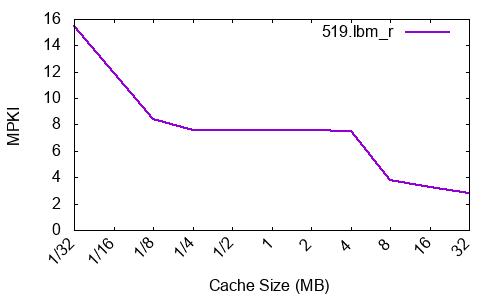}
  \caption{lbm}
\end{subfigure}\hfill
\begin{subfigure}{0.5\columnwidth}
	\includegraphics[width=\textwidth]{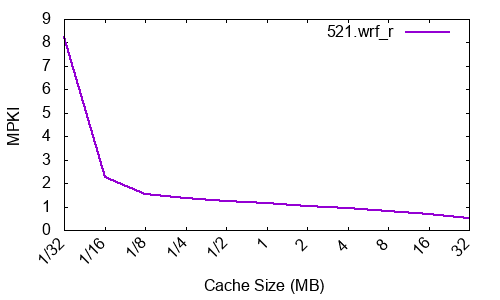}
  \caption{wrf}
\end{subfigure}\hfill
\begin{subfigure}{0.5\columnwidth}
	\includegraphics[width=\textwidth]{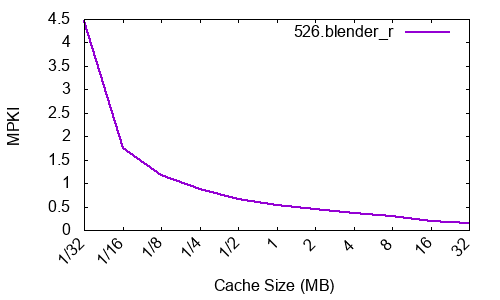}
  \caption{blender}
\end{subfigure}\hfill
\begin{subfigure}{0.5\columnwidth}
	\includegraphics[width=\textwidth]{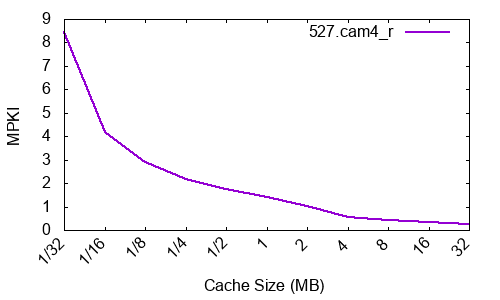}
  \caption{cam4}
\end{subfigure}\hfill
\begin{subfigure}{0.5\columnwidth}
	\includegraphics[width=\textwidth]{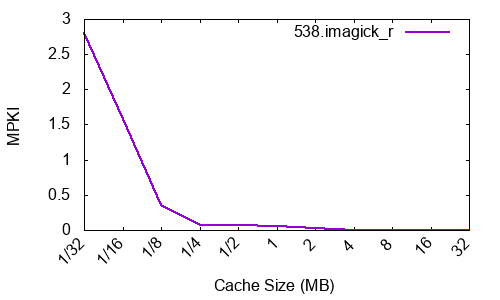}
  \caption{imagick}
\end{subfigure}\hfill
\begin{subfigure}{0.5\columnwidth}
	\includegraphics[width=\textwidth]{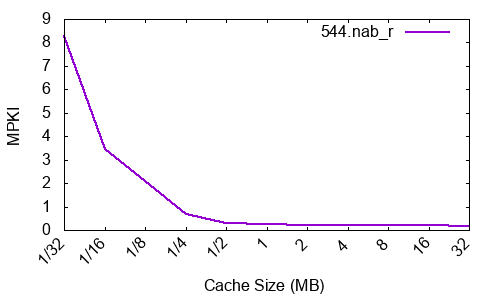}
  \caption{nab}
\end{subfigure}\hfill
\begin{subfigure}{0.5\columnwidth}
	\includegraphics[width=\textwidth]{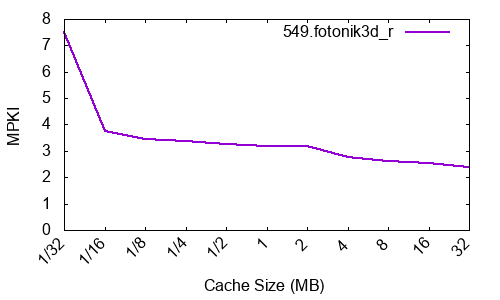}
  \caption{fotonik3d}
\end{subfigure}\hfill
\begin{subfigure}{0.5\columnwidth}
	\includegraphics[width=\textwidth]{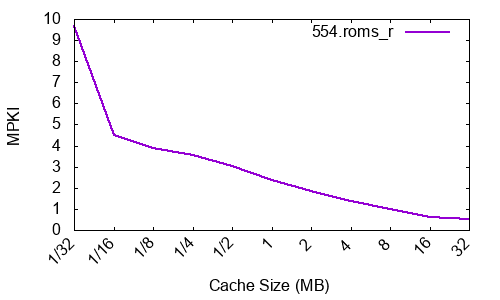}
  \caption{roms}
\end{subfigure}

\caption{Working Set Size}
\label{fig:WorkingSetSize}
\end{figure*}

\par The working set size of an application and its sensitivity to cache capacity can be inferred by examining changes in cache performance of a system with its cache size. For each benchmark, we conduct a cache sensitivity analysis to obtain their working set size. Following the methodology from~\cite{CPU2006_Jaleel}, we modeled a single, shared cache with 64Byte line size and LRU replacement policy, which varied as follows: direct mapped 32KB, 2-way 64KB, 4-way 128KB, 8-way 256KB, and so on till a 1024-way 32MB cache. The experiments are conducted using \textit{dcache}, a functional cache simulator Pintool. Due to dynamic instruction counts in orders of 100 trillion and an effective slowdown incurred by simulation on \textit{dcache}, benchmarks belonging to the FPSpeed suite couldn't be completed and hence are deprecated from the working set size analysis. We consider only one input set for each benchmark.
\par Our results for cache sensivity analysis are presented in Figure~\ref{fig:WorkingSetSize}. We plot cache size in megabytes (MB) on x-axis and misses per kilo instructions (MPKI) on the y-axis. We observe that not all workloads perform well within the range of cache sizes. Based on the working set sizes, we divide the workloads into two group. The first group consists of applications like \textit{povray}, \textit{imagick}, \textit{nab}, and \textit{perlbench} have a limited need for the cache capacity, and can be well executed without the need to regularly refer the main memory. On the contrary, applications like \textit{cactuBSSN}, \textit{lbm}, and \textit{mcf} fail to accomodate within the range of cache sizes, and hence have large working set sizes. The large working sets are often the consequence of the program's algorithm that operates on large amount of data. For example, \textit{cactuBSSN} executes a computational model to employ finite differencing in space using the Einstein equations, while \textit{lbm} simulates an incompressible fluid in 3D. With working set sizes larger than the cache capacity, these applications refer the off-chip memory and hence affect the bandwidth.
\par Figure~\ref{fig:WorkingSetSize} reveals that most workloads exhibit a smooth exponential decrease in the MPKI value as the cache size increases. However, the suite comprises of some workloads where incrementally increasing cache size gives no significant improvements in cache performance, until a point of saturation is reached. At this step, a sudden drop in the MPKI is observed. Such behavior is evident in applications like \textit{bwaves} and \textit{lbm}, and signifies the working set of the workload. Benchamrks like \textit{xalancbmk}, \textit{nab}, \textit{fotonik3d}, and \textit{lbm} illustrate multiple such points, implying that they have multiple working set sizes. Most workloads suffer from cache misses even with a reasonable 32MB cache size, implying that memory hierarchy research, for both on-chip and off-chip components will remain important for these workloads.

%% file: MemoryConsumption.tex
\captionsetup[sub]{font=scriptsize}
\begin{figure*}[hbtp]
\vspace*{-30 pt}
\begin{subfigure}{0.5\columnwidth}
  \includegraphics[width=\textwidth]{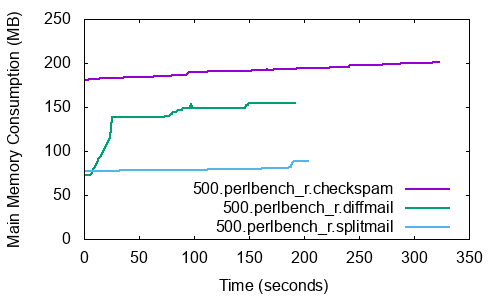}
  \caption{500.perlbench\_r}
\end{subfigure}\hfill
\begin{subfigure}{0.5\columnwidth}
  \includegraphics[width=\textwidth]{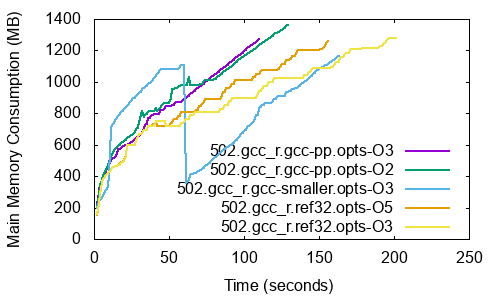}
  \caption{502.gcc\_r}
\end{subfigure}\hfill
\begin{subfigure}{0.5\columnwidth}
  \includegraphics[width=\textwidth]{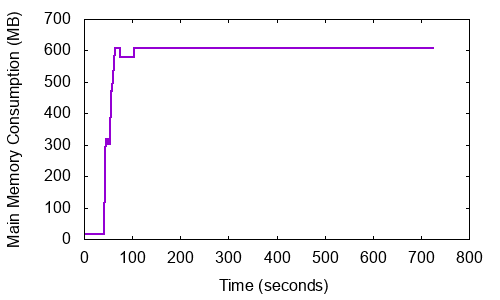}
  \caption{505.mcf\_r}
\end{subfigure}\hfill
\begin{subfigure}{0.5\columnwidth}
  \includegraphics[width=\textwidth]{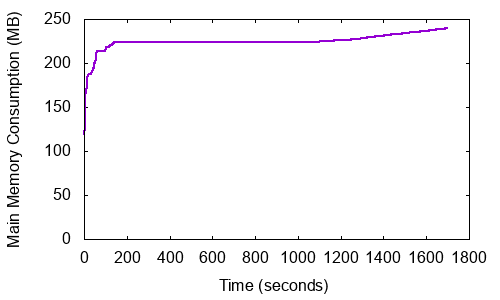}
  \caption{520.omnetpp\_r}
\end{subfigure}\hfill
\begin{subfigure}{0.5\columnwidth}
  \includegraphics[width=\textwidth]{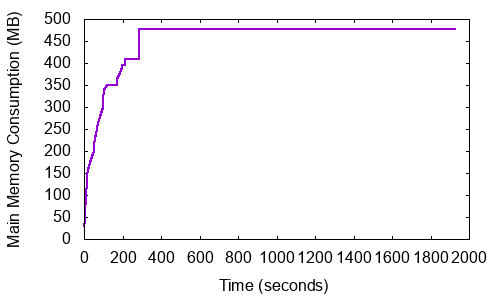}
  \caption{523.xalancbmk\_r}
\end{subfigure}\hfill
\begin{subfigure}{0.5\columnwidth}
  \includegraphics[width=\textwidth]{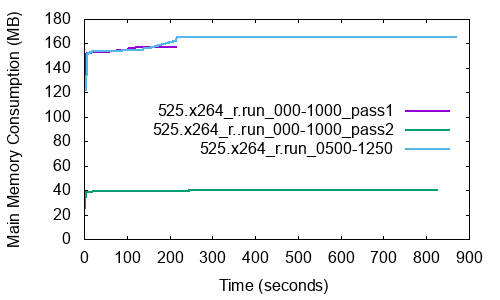}
  \caption{525.x264\_r.}
\end{subfigure}\hfill
\begin{subfigure}{0.5\columnwidth}
  \includegraphics[width=\textwidth]{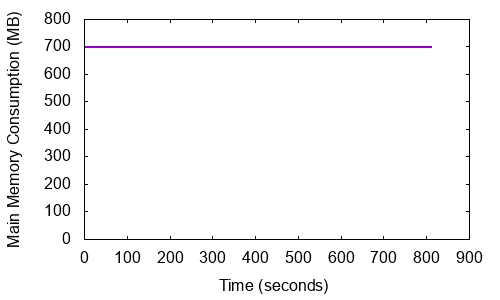}
  \caption{531.deepsjeng\_r}
\end{subfigure}\hfill
\begin{subfigure}{0.5\columnwidth}
  \includegraphics[width=\textwidth]{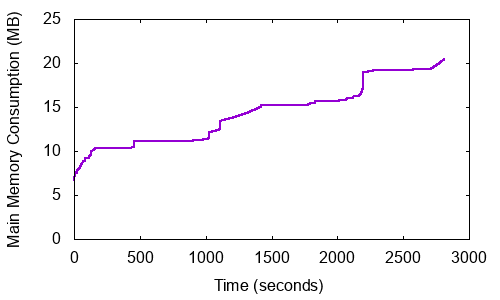}
  \caption{541.leela\_r}
\end{subfigure}\hfill
\begin{subfigure}{0.5\columnwidth}
  \includegraphics[width=\textwidth]{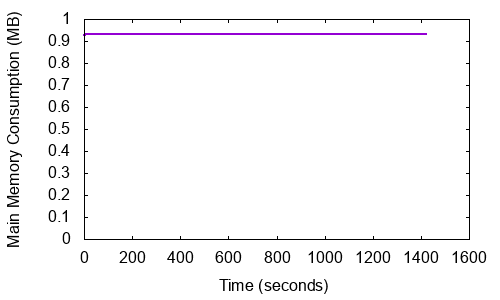}
  \caption{548.exchange2\_r}
\end{subfigure}\hfill
\begin{subfigure}{0.5\columnwidth}
  \includegraphics[width=\textwidth]{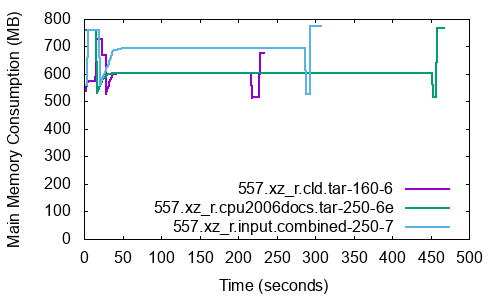}
  \caption{557.xz\_r.}
\end{subfigure}\hfill
\begin{subfigure}{0.5\columnwidth}
  \includegraphics[width=\textwidth]{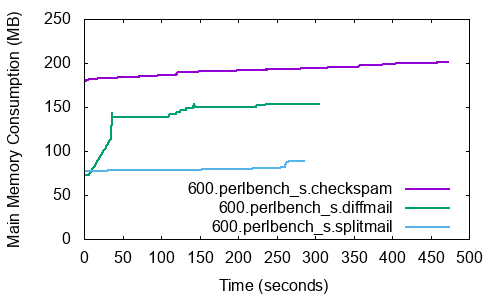}
  \caption{600.perlbench\_s.}
\end{subfigure}\hfill
\begin{subfigure}{0.5\columnwidth}
  \includegraphics[width=\textwidth]{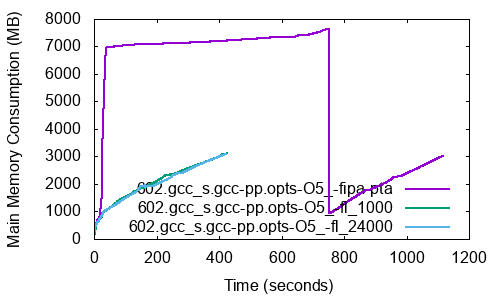}
  \caption{602.gcc\_s.}
\end{subfigure}\hfill
\begin{subfigure}{0.5\columnwidth}
  \includegraphics[width=\textwidth]{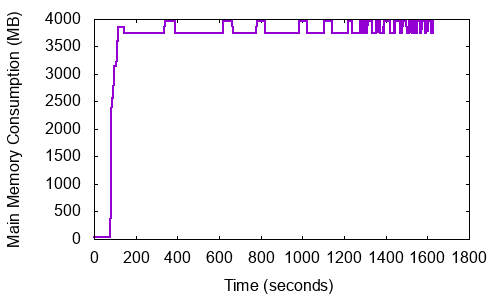}
  \caption{605.mcf\_s}
\end{subfigure}\hfill
\begin{subfigure}{0.5\columnwidth}
  \includegraphics[width=\textwidth]{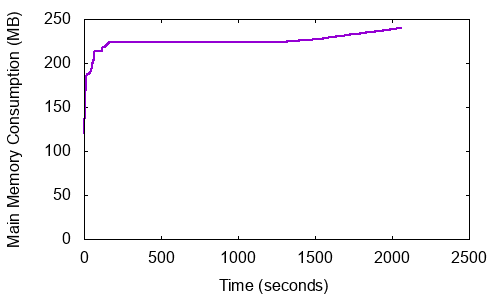}
  \caption{620.omnetpp\_s}
\end{subfigure}\hfill
\begin{subfigure}{0.5\columnwidth}
  \includegraphics[width=\textwidth]{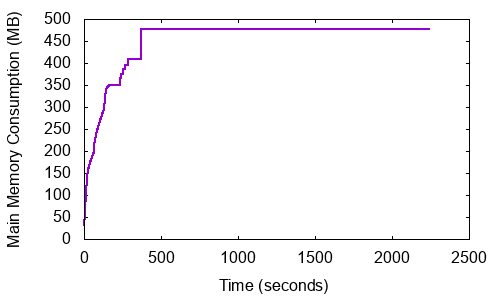}
  \caption{623.xalancbmk\_s}
\end{subfigure}\hfill
\begin{subfigure}{0.5\columnwidth}
  \includegraphics[width=\textwidth]{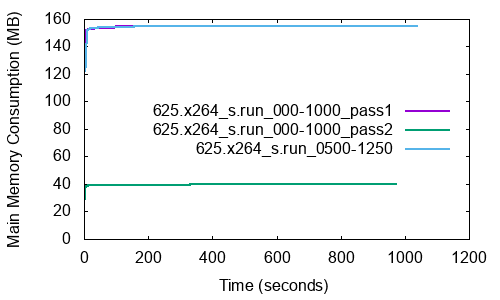}
  \caption{625.x264\_s}
\end{subfigure}\
\begin{subfigure}{0.5\columnwidth}
  \includegraphics[width=\textwidth]{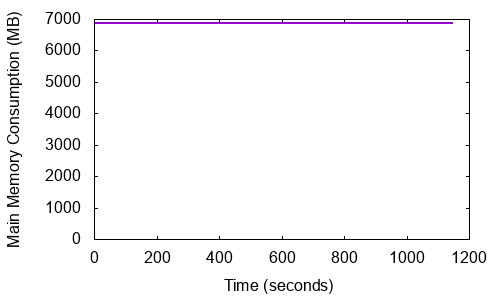}
  \caption{631.deepsjeng\_s}
\end{subfigure}\hfill
\begin{subfigure}{0.5\columnwidth}
  \includegraphics[width=\textwidth]{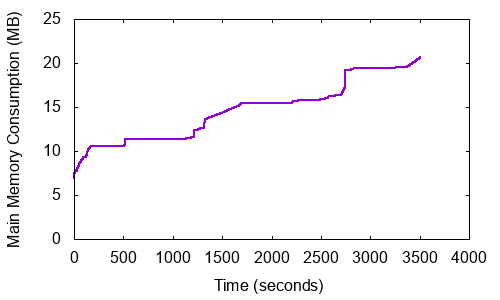}
  \caption{641.leela\_s}
\end{subfigure}\hfill
\begin{subfigure}{0.5\columnwidth}
  \includegraphics[width=\textwidth]{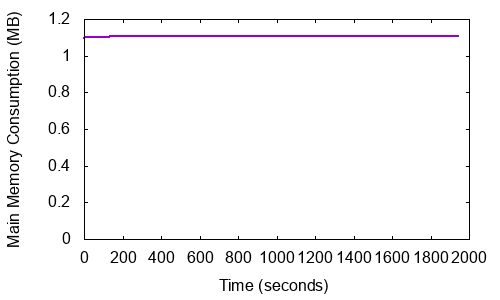}
  \caption{648.exchange2\_s}
\end{subfigure}\hfill
\begin{subfigure}{0.5\columnwidth}
  \includegraphics[width=\textwidth]{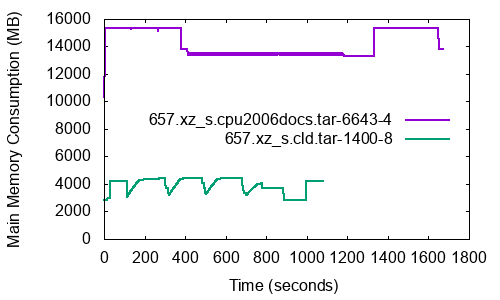}
  \caption{657.xz\_s}
\end{subfigure}\hfill
\begin{subfigure}{0.5\columnwidth}
  \includegraphics[width=\textwidth]{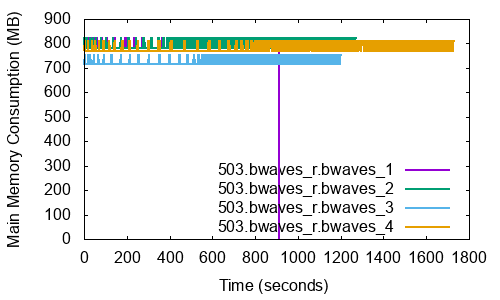}
  \caption{503.bwaves\_r}
\end{subfigure}\hfill
\begin{subfigure}{0.5\columnwidth}
  \includegraphics[width=\textwidth]{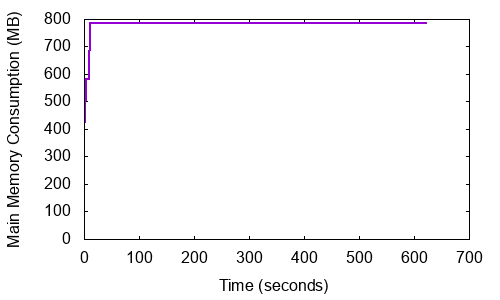}
  \caption{507.cactuBSSN\_r}
\end{subfigure}\hfill
\begin{subfigure}{0.5\columnwidth}
  \includegraphics[width=\textwidth]{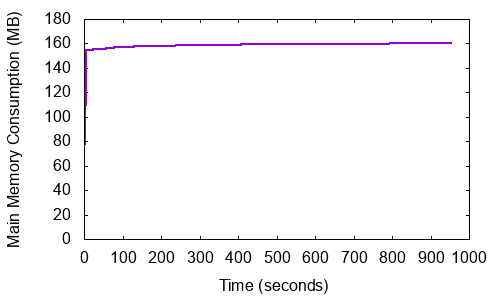}
  \caption{508.namd\_r}
\end{subfigure}\hfill
\begin{subfigure}{0.5\columnwidth}
  \includegraphics[width=\textwidth]{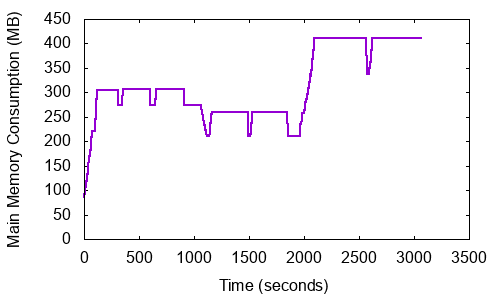}
  \caption{510.parest\_r}
\end{subfigure}\hfill
\begin{subfigure}{0.5\columnwidth}
  \includegraphics[width=\textwidth]{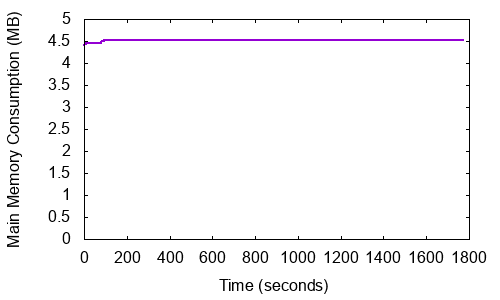}
  \caption{511.povray\_r}
\end{subfigure}\hfill
\begin{subfigure}{0.5\columnwidth}
  \includegraphics[width=\textwidth]{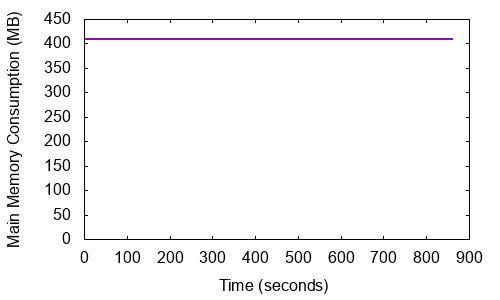}
  \caption{519.lbm\_r}
\end{subfigure}\hfill
\begin{subfigure}{0.5\columnwidth}
  \includegraphics[width=\textwidth]{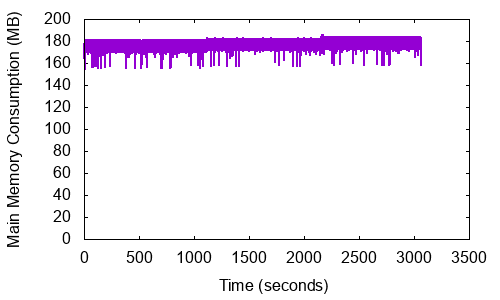}
  \caption{521.wrf\_r}
\end{subfigure}\hfill
\begin{subfigure}{0.5\columnwidth}
  \includegraphics[width=\textwidth]{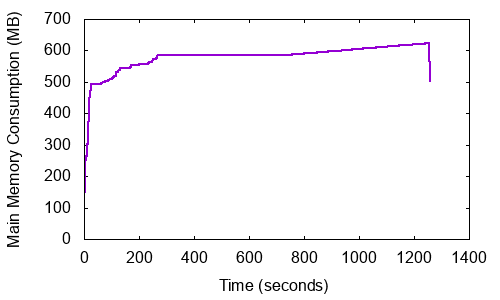}
  \caption{526.blender\_r}
\end{subfigure}
\caption{Main Memory Footprint}
\label{fig:Fig/MemoryConsumption_1}
\end{figure*}

\captionsetup[sub]{font=scriptsize}
\begin{figure*}[hbtp]
\vspace*{-30 pt}
\begin{subfigure}{0.5\columnwidth}
  \includegraphics[width=\textwidth]{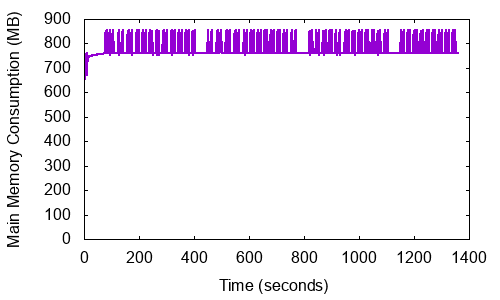}
  \caption{527.cam4\_r}
\end{subfigure}\hfill
\begin{subfigure}{0.5\columnwidth}
  \includegraphics[width=\textwidth]{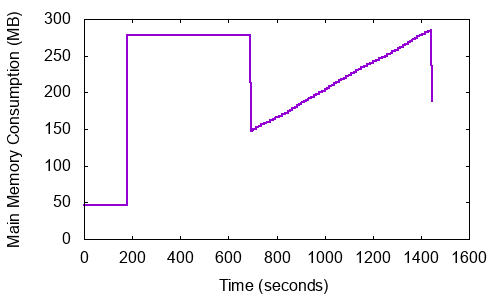}
  \caption{538.imagick\_r}
\end{subfigure}\hfill
\begin{subfigure}{0.5\columnwidth}
  \includegraphics[width=\textwidth]{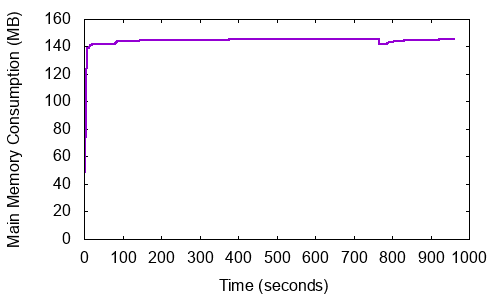}
  \caption{544.nab\_r}
\end{subfigure}\hfill
\begin{subfigure}{0.5\columnwidth}
  \includegraphics[width=\textwidth]{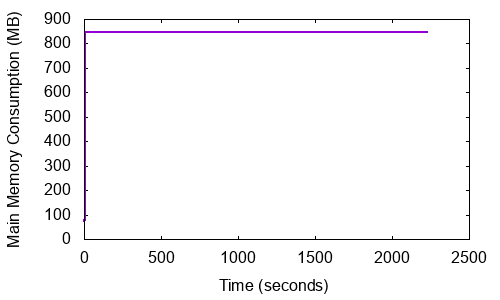}
  \caption{549.fotonik3d\_r}
\end{subfigure}\hfill
\begin{subfigure}{0.5\columnwidth}
  \includegraphics[width=\textwidth]{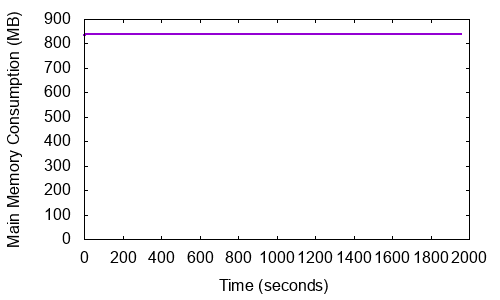}
  \caption{554.roms\_r}
\end{subfigure}\hfill
\begin{subfigure}{0.5\columnwidth}
  \includegraphics[width=\textwidth]{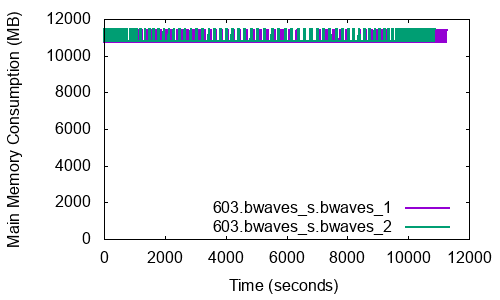}
  \caption{603.bwaves\_s.}
\end{subfigure}\hfill
\begin{subfigure}{0.5\columnwidth}
  \includegraphics[width=\textwidth]{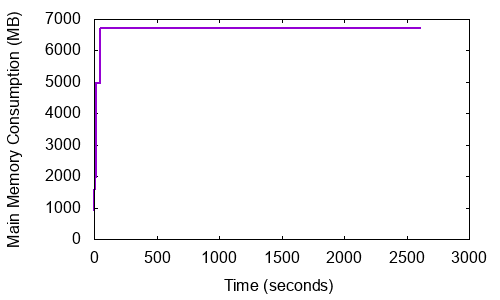}
  \caption{607.cactuBSSN\_s}
\end{subfigure}\hfill
\begin{subfigure}{0.5\columnwidth}
  \includegraphics[width=\textwidth]{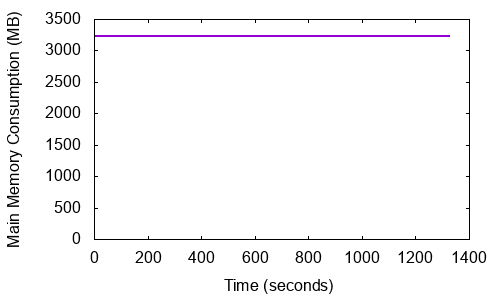}
  \caption{619.lbm\_s}
\end{subfigure}\hfill
\begin{subfigure}{0.5\columnwidth}
  \includegraphics[width=\textwidth]{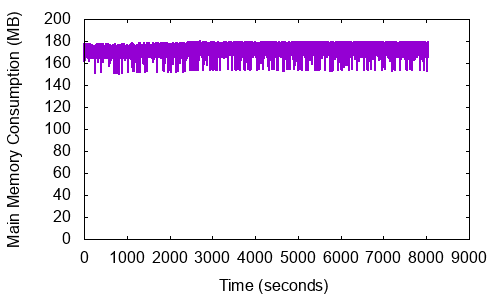}
  \caption{621.wrf\_s}
\end{subfigure}\hfill
\begin{subfigure}{0.5\columnwidth}
  \includegraphics[width=\textwidth]{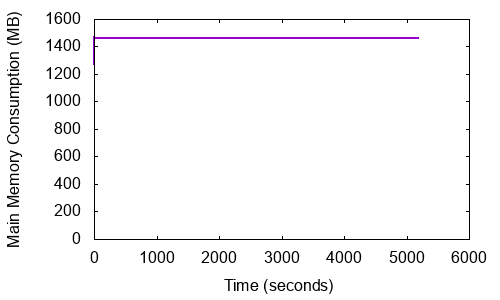}
  \caption{628.pop2\_s}
\end{subfigure}\hfill
\begin{subfigure}{0.5\columnwidth}
  \includegraphics[width=\textwidth]{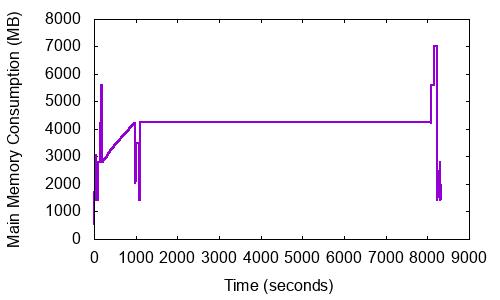}
  \caption{638.imagick\_s}
\end{subfigure}\hfill
\begin{subfigure}{0.5\columnwidth}
  \includegraphics[width=\textwidth]{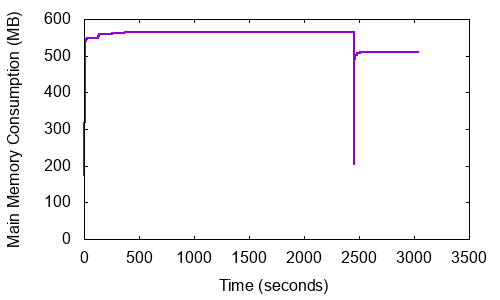}
  \caption{644.nab\_s}
\end{subfigure}\hfill
\begin{subfigure}{0.5\columnwidth}
  \includegraphics[width=\textwidth]{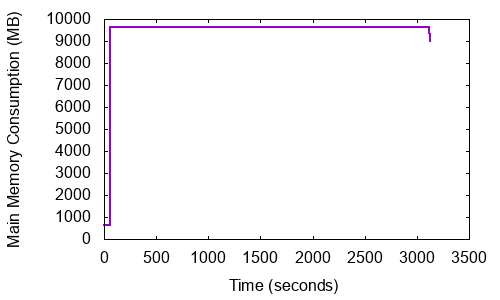}
  \caption{649.fotonik3d\_s}
\end{subfigure}\hfill
\begin{subfigure}{0.5\columnwidth}
  \includegraphics[width=\textwidth]{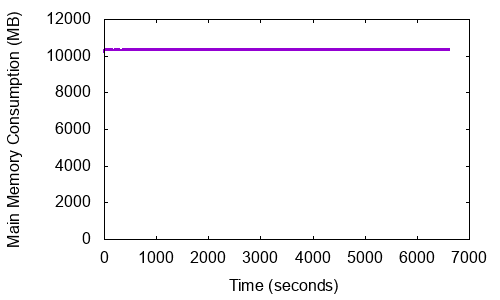}
  \caption{654.roms\_s}
\end{subfigure}
  \vspace*{-20 pt}
\caption{Main Memory Footprint}
\label{fig:Fig/MemoryConsumption_2}
\end{figure*}

\subsection{Memory Footprint}
\label{section:MemoryFootprint}
\par SPEC CPU2006 had a target memory footprint of 900MB for the benchmarks~\cite{CPU2006_WSS,CPU2006_MemoryFootprint}. Since then, the memory size has tremendously increased. We observe the \textit{Resident Set Size (RSS)}, the amount of memory allocated to a process in the main memory, sampled every second, using the Linux \textit{ps} utility. RSS does not include swapped out memory. However, it does include memory from shared libraries as long as the pages from those libraries are actually in memory. A large RSS on an active system means that the process touches a lot of memory locations.
\par Figures~\ref{fig:Fig/MemoryConsumption_1},\ref{fig:Fig/MemoryConsumption_2} plot the time-varying main memory consumption in MB, and indicates that all of the Rate benchmarks, both integer and floating point, still have main memory consumption well below 900MB. However, Speed workloads have large RSS, with peak consumption as high as 16 GB. On average, Speed benchmarks have $\sim$10$\times$ more memory footprint than their corresponding Rate ones. Floating point benchmark suite have memory consumption of $\sim$3$\times$ more than the integer suite. Based on the average footprint throughout the execution, we order the benchmarks from extremely low to extremely high memory consumption. Benchmarks \textit{exchange2}, \textit{povray}, \textit{leela}, \textit{namd}, \textit{wrf}, \textit{nab}, and \textit{xalancbmk} have low RSS values, which indicates negligible access to the main memory. Therefore, these benchmarks are expected have low working set sizes, which is evident from Section~\ref{section:WorkingSet} results. On the contrary, \textit{bwaves\_s}, \textit{roms\_s}, \textit{fotonik3d\_s}, \textit{cactuBSSN\_s} and \textit{xz\_s} exhibit extremely large memory footprints. Furthermore, we observe that $\sim$90\% of the workloads have main memory consumption below 5 GB, resulting in an average memory footprint of 1.82 GB.

%% file: MemoryBandwidth.tex
\subsection{Memory Bandwidth}
\label{section:MemoryBandwidth}
\par
\begin{figure*}[hbtp]
\vspace*{-30 pt}
\begin{subfigure}{0.5\columnwidth}
	\includegraphics[width=\textwidth]{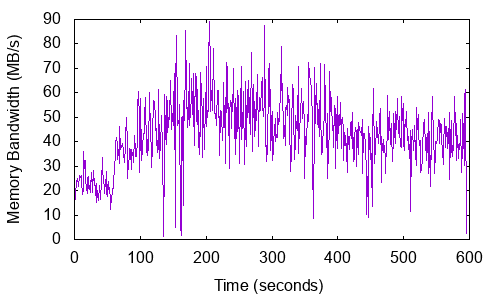}
  \caption{500.perlbench\_r.checkspam}
\end{subfigure}\hfill
\begin{subfigure}{0.5\columnwidth}
	\includegraphics[width=\textwidth]{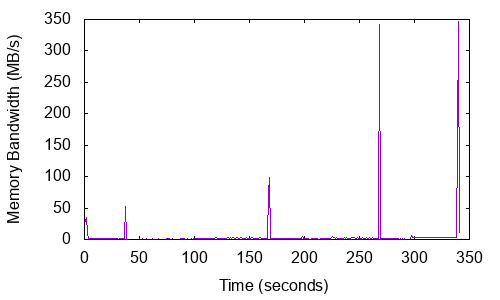}
  \caption{500.perlbench\_r.diffmail}
\end{subfigure}\hfill
\begin{subfigure}{0.5\columnwidth}
	\includegraphics[width=\textwidth]{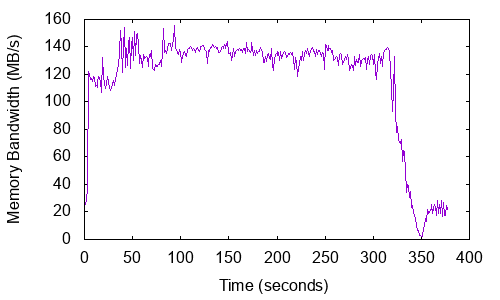}
  \caption{500.perlbench\_r.splitmail}
\end{subfigure}\hfill
\begin{subfigure}{0.5\columnwidth}
	\includegraphics[width=\textwidth]{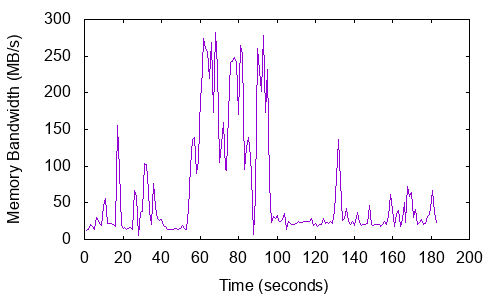}
  \caption{502.gcc\_r.gcc-pp.opts-O3}
\end{subfigure}\hfill
\begin{subfigure}{0.5\columnwidth}
	\includegraphics[width=\textwidth]{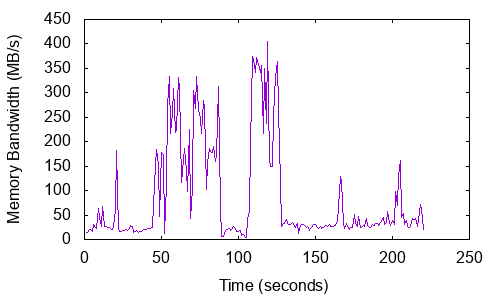}
  \caption{502.gcc\_r.gcc-pp.opts-O2}
\end{subfigure}\hfill
\begin{subfigure}{0.5\columnwidth}
	\includegraphics[width=\textwidth]{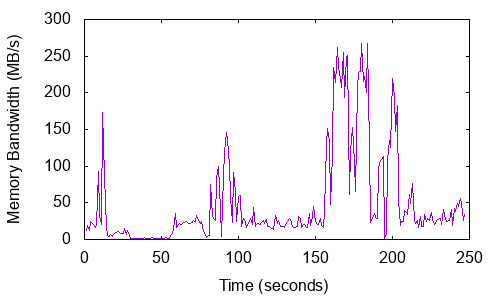}
  \caption{502.gcc\_r.gcc-smaller.opts-O3}
\end{subfigure}\hfill
\begin{subfigure}{0.5\columnwidth}
	\includegraphics[width=\textwidth]{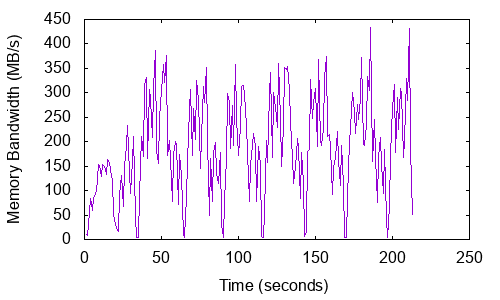}
  \caption{502.gcc\_r.ref32.opts-O5}
\end{subfigure}\hfill
\begin{subfigure}{0.5\columnwidth}
	\includegraphics[width=\textwidth]{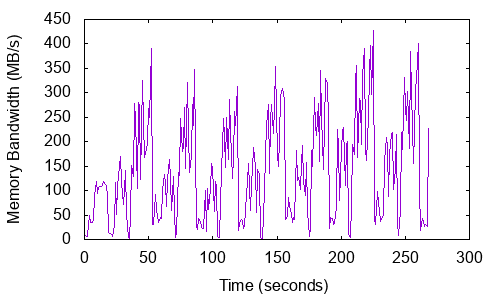}
  \caption{502.gcc\_r.ref32.opts-O3}
\end{subfigure}\hfill
\begin{subfigure}{0.5\columnwidth}
	\includegraphics[width=\textwidth]{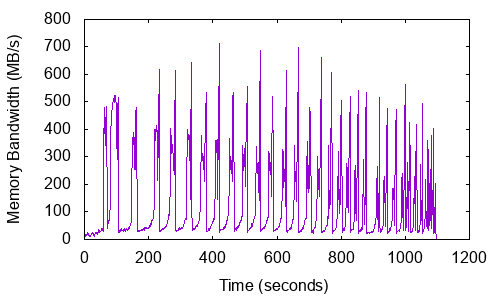}
  \caption{505.mcf\_r}
\end{subfigure}\hfill
\begin{subfigure}{0.5\columnwidth}
	\includegraphics[width=\textwidth]{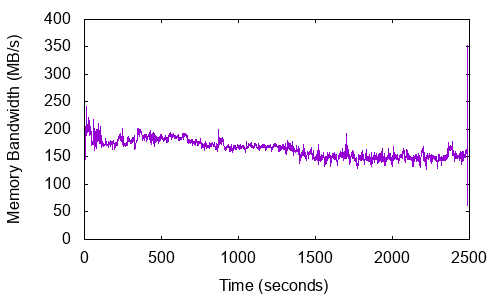}
  \caption{520.omnetpp\_r}
\end{subfigure}\hfill
\begin{subfigure}{0.5\columnwidth}
	\includegraphics[width=\textwidth]{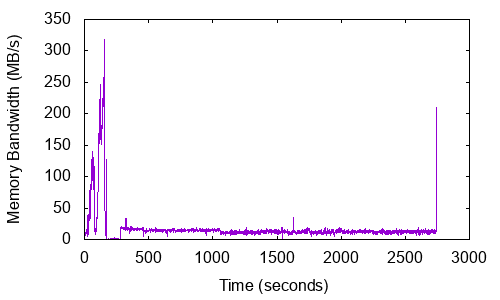}
  \caption{523.xalancbmk\_r}
\end{subfigure}\hfill
\begin{subfigure}{0.5\columnwidth}
	\includegraphics[width=\textwidth]{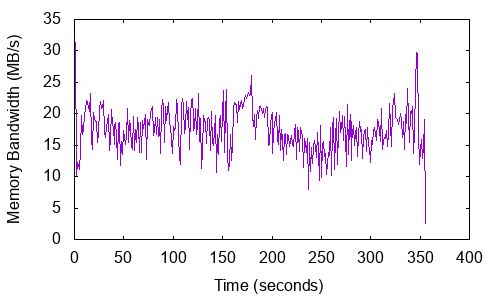}
  \caption{525.x264\_r.run\_000-1000\_pass1}
\end{subfigure}\hfill
\begin{subfigure}{0.5\columnwidth}
	\includegraphics[width=\textwidth]{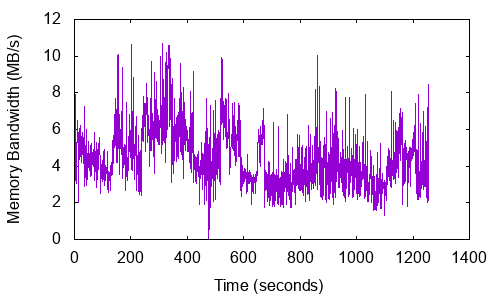}
  \caption{525.x264\_r.run\_000-1000\_pass2}
\end{subfigure}\hfill
\begin{subfigure}{0.5\columnwidth}
	\includegraphics[width=\textwidth]{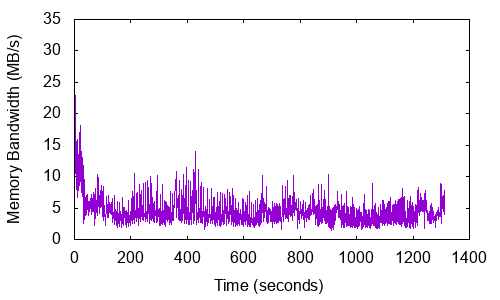}
  \caption{525.x264\_r.run\_0500-1250}
\end{subfigure}\hfill
\begin{subfigure}{0.5\columnwidth}
	\includegraphics[width=\textwidth]{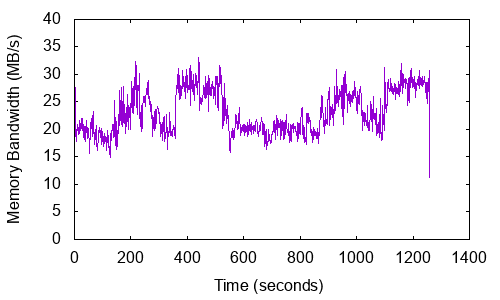}
  \caption{531.deepsjeng\_r}
\end{subfigure}\hfill
\begin{subfigure}{0.5\columnwidth}
	\includegraphics[width=\textwidth]{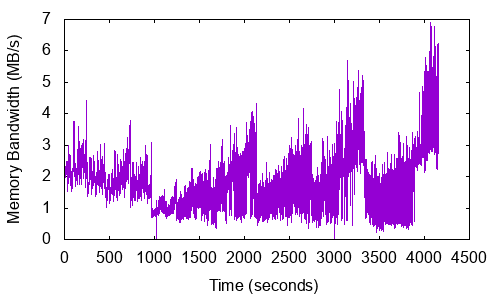}
  \caption{541.leela\_r}
\end{subfigure}\hfill
\begin{subfigure}{0.5\columnwidth}
	\includegraphics[width=\textwidth]{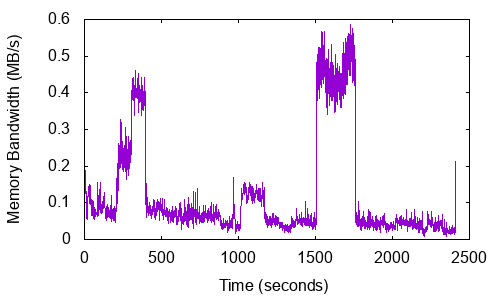}
  \caption{548.exchange2\_r}
\end{subfigure}\hfill
\begin{subfigure}{0.5\columnwidth}
	\includegraphics[width=\textwidth]{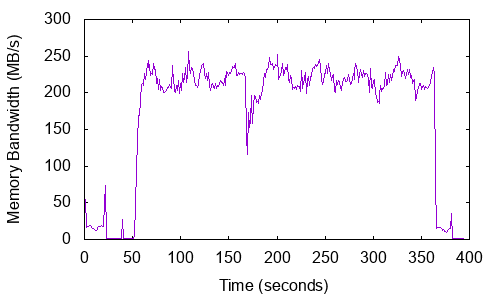}
  \caption{557.xz\_r.cld.tar-160-6}
\end{subfigure}\hfill
\begin{subfigure}{0.5\columnwidth}
	\includegraphics[width=\textwidth]{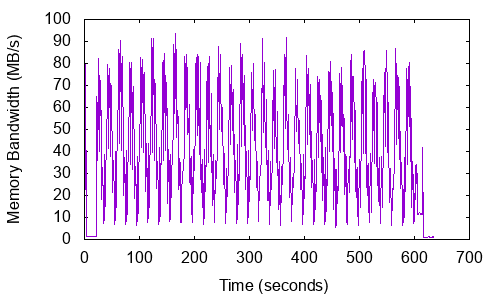}
  \caption{557.xz\_r.cpu2006docs.tar-250-6e}
\end{subfigure}\hfill
\begin{subfigure}{0.5\columnwidth}
	\includegraphics[width=\textwidth]{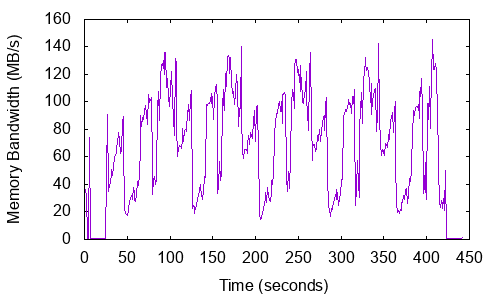}
  \caption{557.xz\_r.input.combined-250-7}
\end{subfigure}\hfill
\begin{subfigure}{0.5\columnwidth}
	\includegraphics[width=\textwidth]{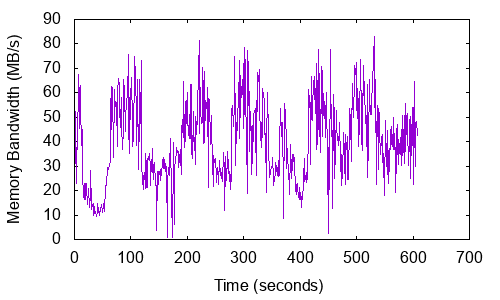}
  \caption{600.perlbench\_s.checkspam}
\end{subfigure}\hfill
\begin{subfigure}{0.5\columnwidth}
	\includegraphics[width=\textwidth]{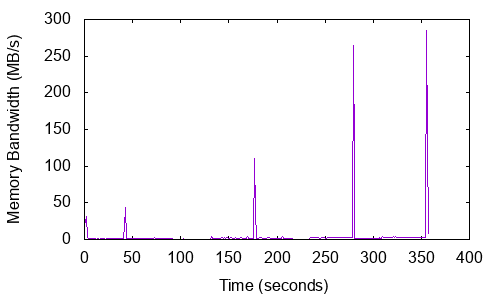}
  \caption{600.perlbench\_s.diffmail}
\end{subfigure}\hfill
\begin{subfigure}{0.5\columnwidth}
	\includegraphics[width=\textwidth]{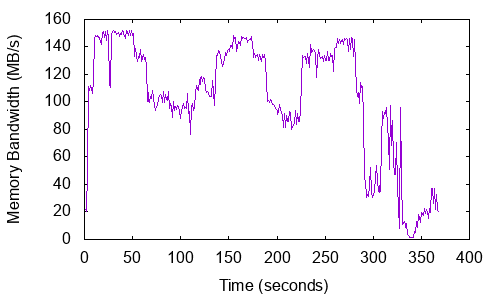}
  \caption{600.perlbench\_s.splitmail}
\end{subfigure}\hfill
\begin{subfigure}{0.5\columnwidth}
	\includegraphics[width=\textwidth]{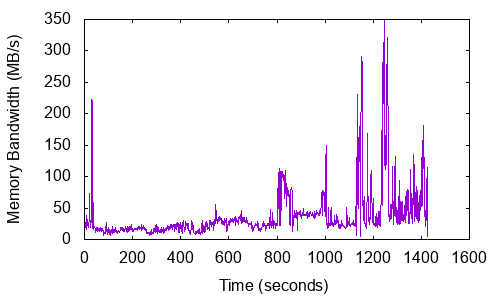}
  \caption{602.gcc\_s.gcc-pp.opts-O5\_-fipa-pta}
\end{subfigure}\hfill
\begin{subfigure}{0.5\columnwidth}
	\includegraphics[width=\textwidth]{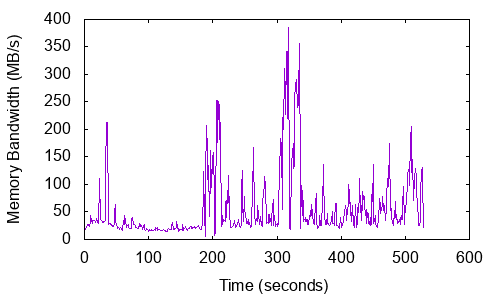}
  \caption{602.gcc\_s.gcc-pp.opts-O5\_-fl\_1000}
\end{subfigure}\hfill
\begin{subfigure}{0.5\columnwidth}
	\includegraphics[width=\textwidth]{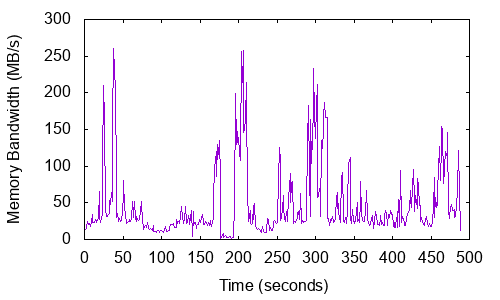}
  \caption{602.gcc\_s.gcc-pp.opts-O5\_-fl\_24000}
\end{subfigure}\hfill
\begin{subfigure}{0.5\columnwidth}
	\includegraphics[width=\textwidth]{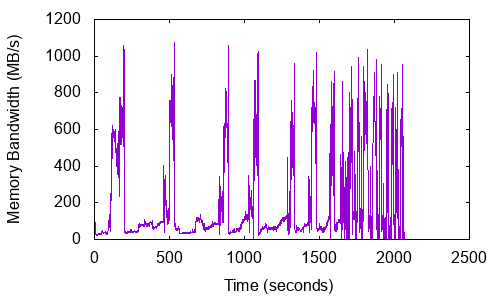}
  \caption{605.mcf\_s}
\end{subfigure}\hfill
\begin{subfigure}{0.5\columnwidth}
	\includegraphics[width=\textwidth]{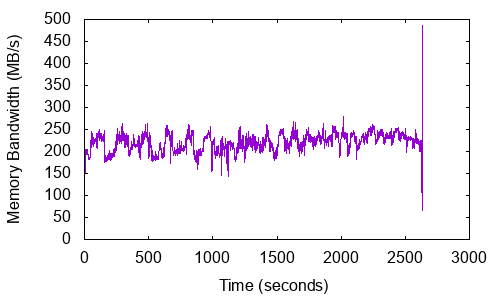}
  \caption{620.omnetpp\_s}
\end{subfigure}\hfill
\caption{Memory Bandwidth}
\label{fig:Fig/MemoryBandwidth_1}
\end{figure*}

\begin{figure*}[hbtp]
\vspace*{-30 pt}
\begin{subfigure}{0.5\columnwidth}
	\includegraphics[width=\textwidth]{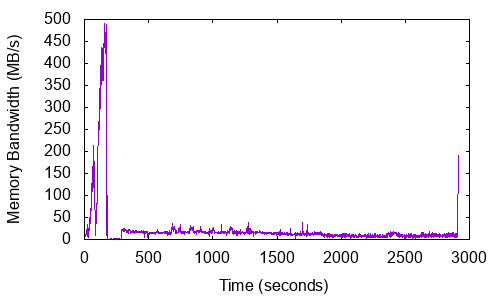}
  \caption{623.xalancbmk\_s}
\end{subfigure}\hfill
\begin{subfigure}{0.5\columnwidth}
	\includegraphics[width=\textwidth]{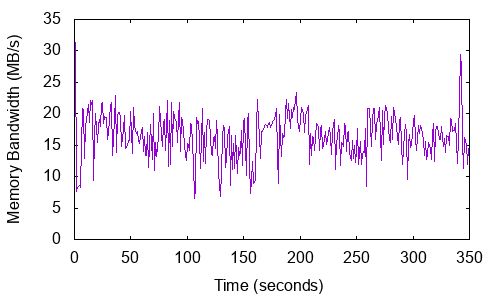}
  \caption{625.x264\_s.run\_000-1000\_pass1}
\end{subfigure}\hfill
\begin{subfigure}{0.5\columnwidth}
	\includegraphics[width=\textwidth]{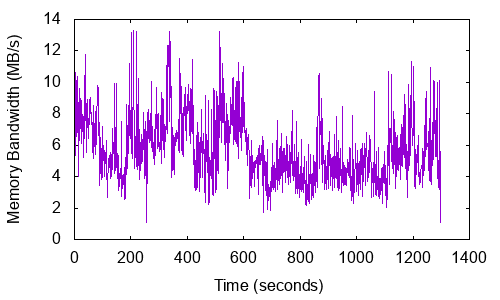}
  \caption{625.x264\_s.run\_000-1000\_pass2}
\end{subfigure}\hfill
\begin{subfigure}{0.5\columnwidth}
	\includegraphics[width=\textwidth]{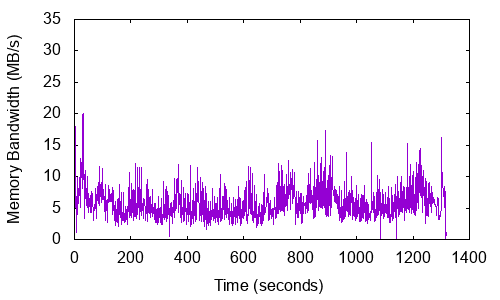}
  \caption{625.x264\_s.run\_0500-1250}
\end{subfigure}\hfill
\begin{subfigure}{0.5\columnwidth}
	\includegraphics[width=\textwidth]{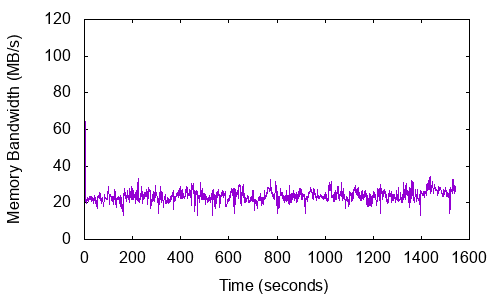}
  \caption{631.deepsjeng\_s}
\end{subfigure}\hfill
\begin{subfigure}{0.5\columnwidth}
	\includegraphics[width=\textwidth]{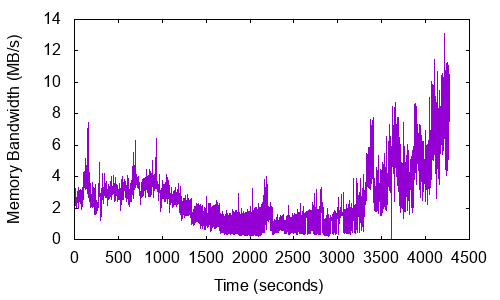}
  \caption{641.leela\_s}
\end{subfigure}\hfill
\begin{subfigure}{0.5\columnwidth}
	\includegraphics[width=\textwidth]{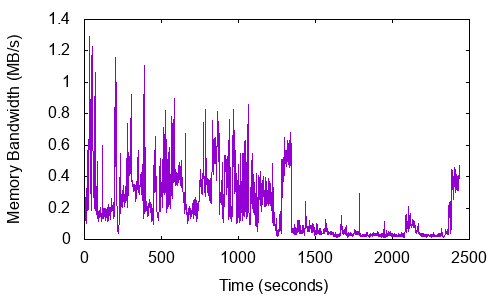}
  \caption{648.exchange2\_s}
\end{subfigure}\hfill
\begin{subfigure}{0.5\columnwidth}
	\includegraphics[width=\textwidth]{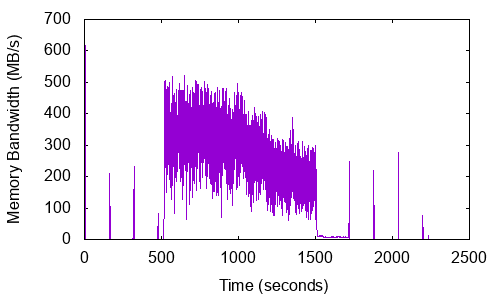}
  \caption{657.xz\_s.cpu2006docs.tar-6643-4}
\end{subfigure}\hfill
\begin{subfigure}{0.5\columnwidth}
	\includegraphics[width=\textwidth]{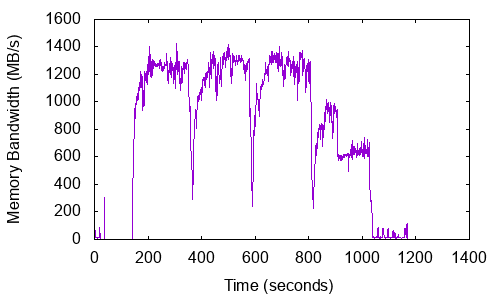}
  \caption{657.xz\_s.cld.tar-1400-8}
\end{subfigure}\hfill
\begin{subfigure}{0.5\columnwidth}
	\includegraphics[width=\textwidth]{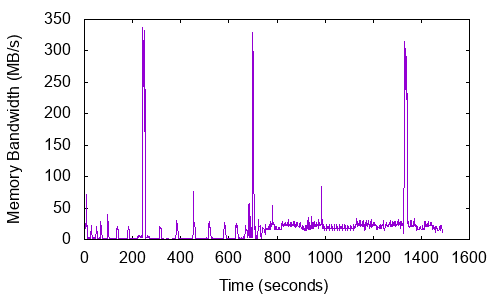}
  \caption{503.bwaves\_r.bwaves\_1}
\end{subfigure}\hfill
\begin{subfigure}{0.5\columnwidth}
	\includegraphics[width=\textwidth]{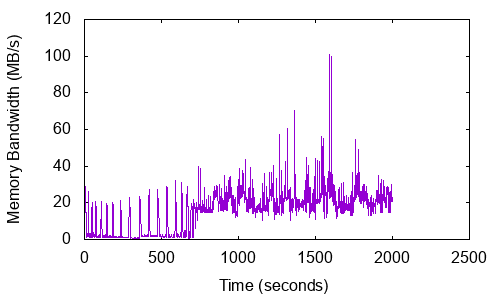}
  \caption{503.bwaves\_r.bwaves\_2}
\end{subfigure}\hfill
\begin{subfigure}{0.5\columnwidth}
	\includegraphics[width=\textwidth]{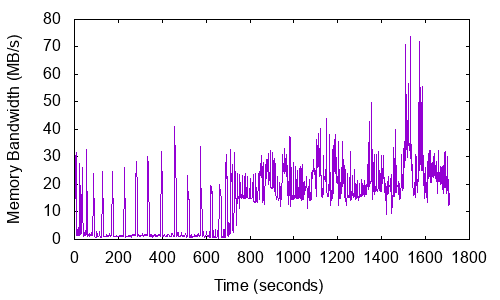}
  \caption{503.bwaves\_r.bwaves\_3}
\end{subfigure}\hfill
\begin{subfigure}{0.5\columnwidth}
	\includegraphics[width=\textwidth]{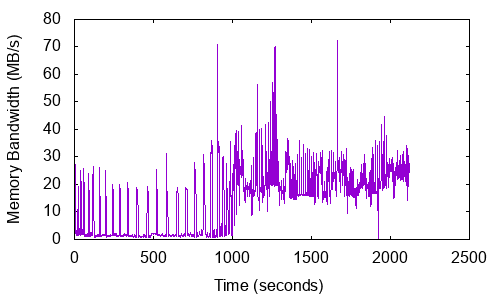}
  \caption{503.bwaves\_r.bwaves\_4}
\end{subfigure}\hfill
\begin{subfigure}{0.5\columnwidth}
	\includegraphics[width=\textwidth]{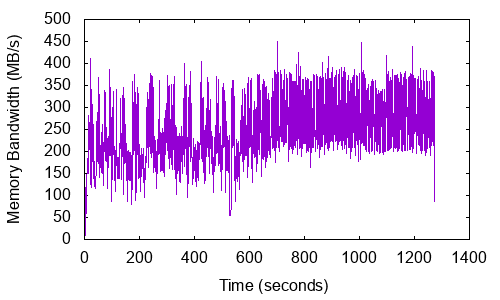}
  \caption{507.cactuBSSN\_r}
\end{subfigure}\hfill
\begin{subfigure}{0.5\columnwidth}
	\includegraphics[width=\textwidth]{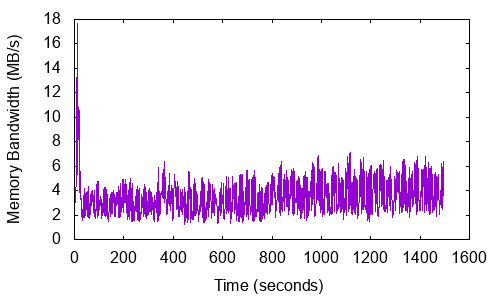}
  \caption{508.namd\_r}
\end{subfigure}\hfill
\begin{subfigure}{0.5\columnwidth}
	\includegraphics[width=\textwidth]{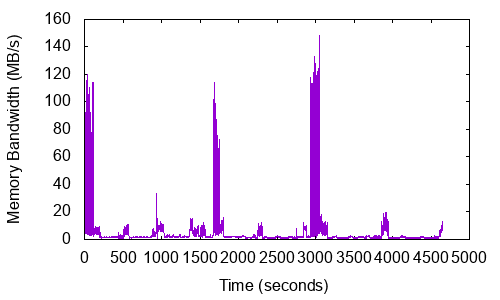}
  \caption{510.parest\_r}
\end{subfigure}\hfill
\begin{subfigure}{0.5\columnwidth}
	\includegraphics[width=\textwidth]{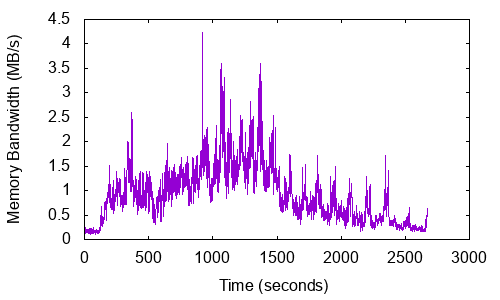}
  \caption{511.povray\_r}
\end{subfigure}\hfill
\begin{subfigure}{0.5\columnwidth}
	\includegraphics[width=\textwidth]{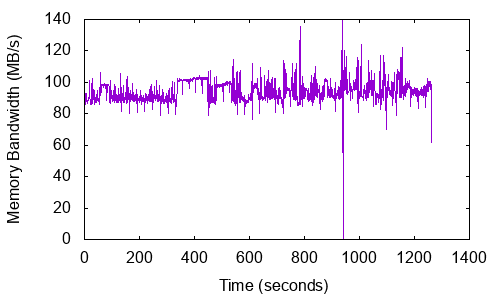}
  \caption{519.lbm\_r}
\end{subfigure}\hfill
\begin{subfigure}{0.5\columnwidth}
	\includegraphics[width=\textwidth]{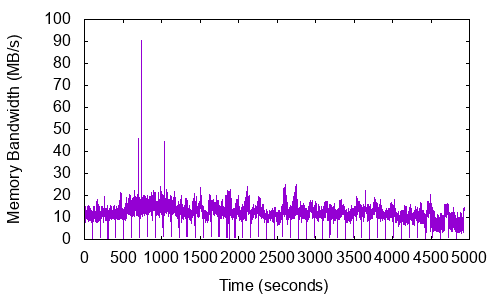}
  \caption{521.wrf\_r}
\end{subfigure}\hfill
\begin{subfigure}{0.5\columnwidth}
	\includegraphics[width=\textwidth]{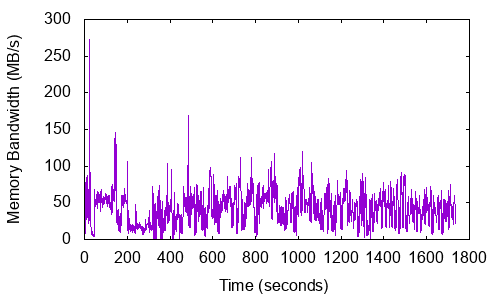}
  \caption{526.blender\_r}
\end{subfigure}\hfill
\begin{subfigure}{0.5\columnwidth}
	\includegraphics[width=\textwidth]{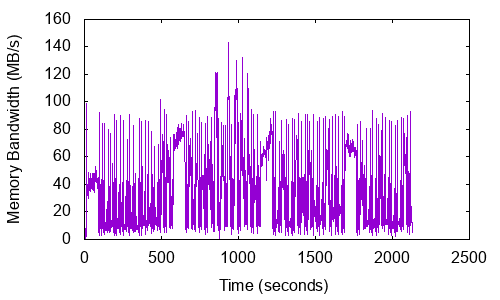}
  \caption{527.cam4\_r}
\end{subfigure}\hfill
\begin{subfigure}{0.5\columnwidth}
	\includegraphics[width=\textwidth]{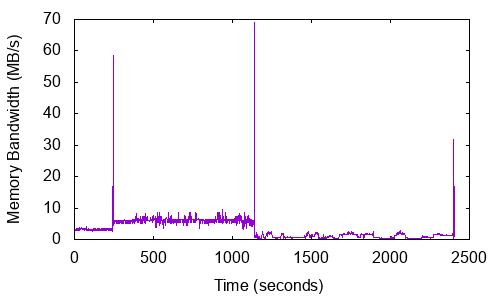}
  \caption{538.imagick\_r}
\end{subfigure}\hfill
\begin{subfigure}{0.5\columnwidth}
	\includegraphics[width=\textwidth]{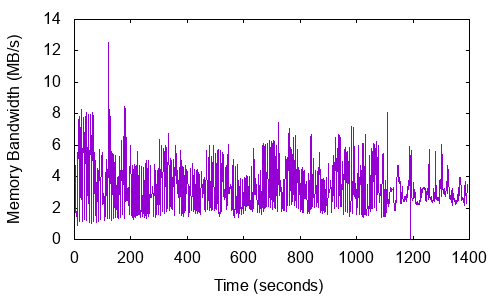}
  \caption{544.nab\_r}
\end{subfigure}\hfill
\begin{subfigure}{0.5\columnwidth}
	\includegraphics[width=\textwidth]{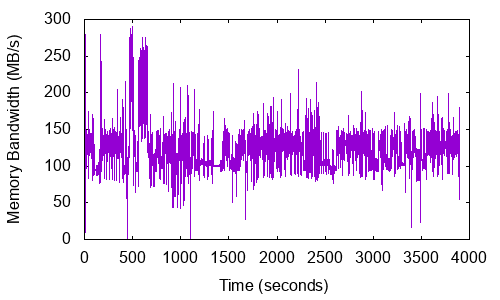}
  \caption{549.fotonik3d\_r}
\end{subfigure}\hfill
\begin{subfigure}{0.5\columnwidth}
	\includegraphics[width=\textwidth]{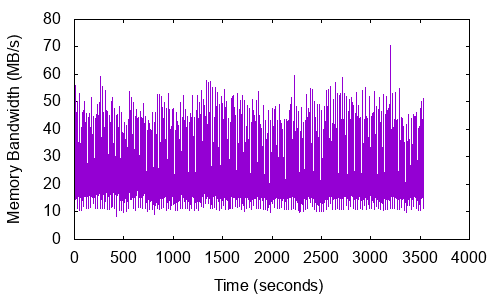}
  \caption{554.roms\_r}
\end{subfigure}\hfill
\begin{subfigure}{0.5\columnwidth}
	\includegraphics[width=\textwidth]{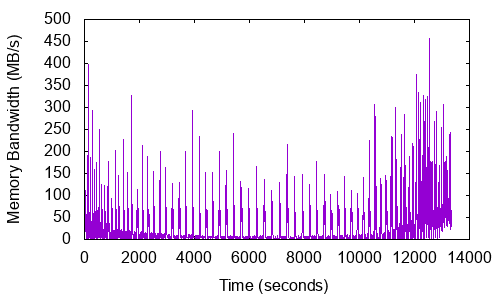}
  \caption{603.bwaves\_s.bwaves\_1}
\end{subfigure}\hfill
\begin{subfigure}{0.5\columnwidth}
	\includegraphics[width=\textwidth]{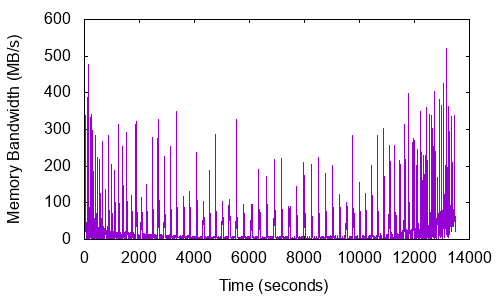}
  \caption{603.bwaves\_s.bwaves\_2}
\end{subfigure}\hfill
\begin{subfigure}{0.5\columnwidth}
	\includegraphics[width=\textwidth]{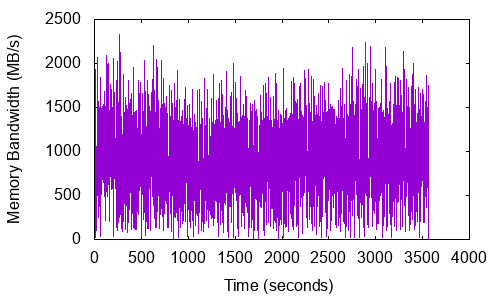}
  \caption{607.cactuBSSN\_s}
\end{subfigure}
\caption{Memory Bandwidth}
\label{fig:Fig/MemoryBandwidth_2}
\end{figure*}

\begin{figure*}[hbtp]
\vspace*{-30 pt}
\begin{subfigure}{0.5\columnwidth}
	\includegraphics[width=\textwidth]{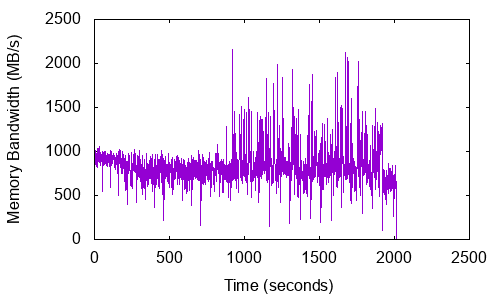}
  \caption{619.lbm\_s}
\end{subfigure}\hfill
\begin{subfigure}{0.5\columnwidth}
	\includegraphics[width=\textwidth]{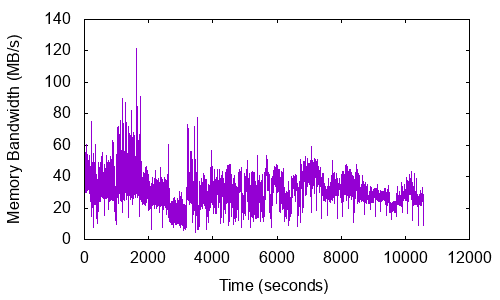}
  \caption{621.wrf\_s}
\end{subfigure}\hfill
\begin{subfigure}{0.5\columnwidth}
	\includegraphics[width=\textwidth]{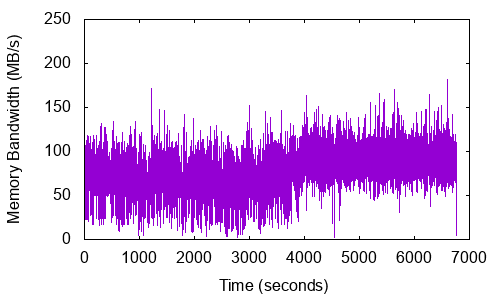}
  \caption{628.pop2\_s}\hfill
\end{subfigure}\hfill
\begin{subfigure}{0.5\columnwidth}
	\includegraphics[width=\textwidth]{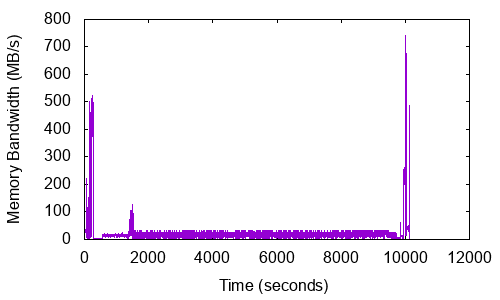}
  \caption{638.imagick\_s}
\end{subfigure}\hfill
\begin{subfigure}{0.5\columnwidth}
	\includegraphics[width=\textwidth]{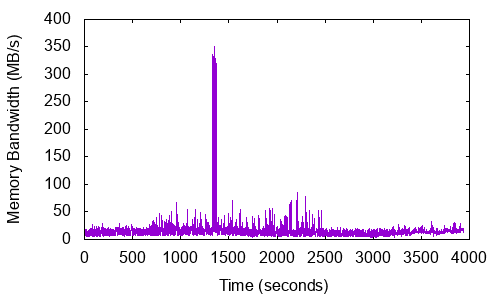}
  \caption{644.nab\_s}
\end{subfigure}\hfill
\begin{subfigure}{0.5\columnwidth}
	\includegraphics[width=\textwidth]{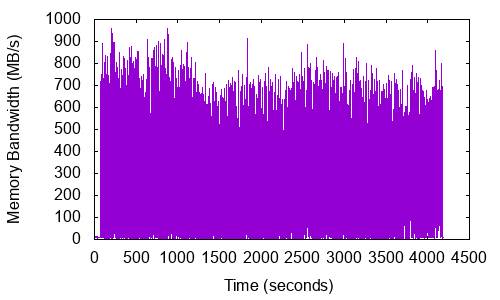}
  \caption{649.fotonik3d\_s}
\end{subfigure}\hfill
\begin{subfigure}{0.5\columnwidth}
	\includegraphics[width=\textwidth]{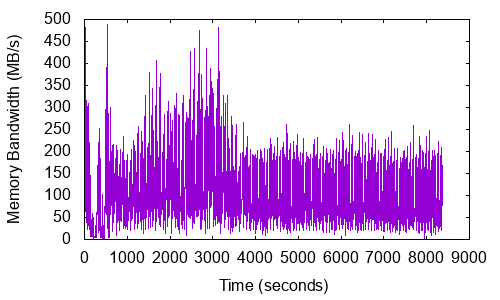}
  \caption{654.roms\_s}
\end{subfigure}
\caption{Memory Bandwidth}
\label{fig:Fig/MemoryBandwidth_3}
\end{figure*}

 Next, we measure the off-chip bandwidth across the SPEC CPU2017 workloads. We collect the hardware events \textit{LLC-load-misses} and \textit{LLC-store-misses} using perf at regular intervals of 1 second, on test system described in Table~\ref{table:CAD_config}. Memory bandwidth is calculated as the product of the total LLC misses per second with the cache line size. Figures~\ref{fig:Fig/MemoryBandwidth_1}-\ref{fig:Fig/MemoryBandwidth_3} plots the time-varying memory bandwidth results in Megabytes per second, for each workload.
\par Our experimental results indicate a large variety in memory bandwidth usage patterns from various benchmarks. CPU2017 consists of workloads with average bandwidth as low as 0.2 MB/s to workloads with peak bandwidth of 2.3 GB/s. \textit{leela}, \textit{exchange2}, \textit{namd}, \textit{povray}, and \textit{nab\_r} have modest bandwidth usage, with consumption within 10 MB/s during the entire execution period. Workloads \textit{parest}, \textit{wrf\_r}, \textit{nab\_s} and \textit{perlbench.diffmail} exhibit low bandwidth usage with short sudden irregular bursts of high data transfer rates. While applications like \textit{xalancbmk} and \textit{imagick} have input sets which fit within on-chip memory, and hence these applications do not refer the off-chip memory after initiation. All the above discussed benchmarks have very little off-chip bandwidth usage. This is in line with the conclusions drawn from Sections~\ref{section:WorkingSet} and~\ref{section:MemoryFootprint}, as these workloads have low working set sizes and hence low memory footprint.  
\par CPU2017 also comprises of many benchmarks with large memory bandwidth utilizations. For example, \textit{cactuBSSN\_s}, and \textit{lbm\_s} have peak bandwidth utilization of 2.3 GB/s (0.9 GB/s on average). Similarly, \textit{mcf}, \textit{xz\_s}, \textit{cactuBSSN\_r}, and \textit{fotonik3d\_s} have also large off-chip traffic, and can be used to test bandwidth optimization techniques.  

\captionsetup{justification=raggedright}
\captionsetup[sub]{font=scriptsize}

\captionsetup{justification=raggedright}
\captionsetup[sub]{font=scriptsize}

\captionsetup[sub]{font=scriptsize}

%% file: NewBenchmarks.tex
\section{New Additions to the SPEC CPU suite}
\label{section:NewBenchmarks}

\par In the current iteration of SPEC CPU, many new benchmarks have been added to cover emerging application domains. In the INT category, artificial intelligence (AI) has been extensively represented by a total of three benchmarks, with \textit{exchange2} being the new addition to the group. CPU2006~\cite{CPU2006} integer benchmarks \textit{h264ref}, \textit{sjeng} and \textit{gobmk} have been renamed to \textit{x264}, \textit{deepsjeng} and \textit{leela} respectively due to changes in their functionality or inputs, while still maintaining the application domain. Additionally, \textit{bzip2} has been replaced by \textit{xz} to represent the general compression domain. \textit{exchange2} (recursive solution generator), the new addition to INT suite, has the lowest percentage of memory instructions and hence, justifiably the lowest memory footprint and lowest bandwidth consumption in the CPU2017 suite. Interestingly, all the three AI benchmarks in the suite have extremely small working set sizes and consequently, low off-chip accesses.
\par In the FP category, eight new benchmarks have been added: \textit{parest}, \textit{blender}, \textit{cam4}, \textit{pop2}, \textit{imagick}, \textit{nab}, \textit{fotonik3d}, and \textit{roms}. Climatology domain has been extensively represented here with three new additions of benchmarks, simulating different components of the NCAR Earth System. \textit{cactusADM} has been changed to \textit{cactuBSSN}. \textit{parest's} implementation relies on \textit{dealII} libraries from CPU2006, which also underlines the \textit{dealII} benchmark.
 In general, Speed versions of these benchmarks are scaled up in order to highly exercise both memory and computation. For example, \textit{xz} achieves this by differing in its data compression levels, \textit{roms} vary its grid size and simulation time steps, while \textit{fotonik3d} alters its problem size, frequencies, time steps, and boundary conditions. At the same time, benchmarks \textit{x264}, \textit{leela} and \textit{exchange2} use almost similar workloads for both Rate and Speed and hence, we discern very similar instruction and memory behavior from them, as depicted throughout the Sections~\ref{section:instructionprofile} and~\ref{section:MemoryBehavior}.

%% file: related.tex
\section{Related Work}\label{section:related_work}
A number of studies have been carried out recently regarding characterization of SPEC CPU2017 workloads, however, to the best of our knowledge, this paper presents the first systematic study of the memory behavior of the SPEC CPU2017 suite. 
\par \textit{SPEC CPU2017 Characterization:} 
\citet{CPU2017_Description} present an overview of CPU2017 suite and discuss its reportable execution. \citet{CPU2017_characterization} use hardware performance counter statistics to characterize SPEC CPU2017 applications with respect to several metrics such as instruction distribution, execution performance, branch and cache behaviors. They also utilize Principal Components Analysis~\cite{PCA} and hierarchical clustering to identify subsets of the suite. Similarly, \citet{CPU2017_WaitofaDecade} characterize the CPU2017 benchmarks using \textit{perf}, and leverage statistical techniques to identify cross application redundancies and propose subsets of the entire suite, by classifying multiple benchmarks with similar behaviors into a single subset. Further, they also provide a detailed evaluation of the representativeness of the subsets. \citet{CPU2017_Alberta} propose the Alberta Workloads for the SPEC CPU2017 benchmark suite hoping to improve the performance evaluation of techniques that rely on any type of learning, for example the formal Feedback-Directed Optimization (FDO). Additionally, in order to ameliorate large simulation times, \citet{CPU2017_Hotregions} analyze the program behavior and consequently propose simulation points~\cite{SimulationPoints} for the suite.
\par \textit{Memory Characterization of Workloads:}
\citet{CPU2006_Jaleel} determined the memory system requirements of workloads from SPEC CPU2000 and CPU2006 using binary instrumentation. \citet{CPU2006_MemoryFootprint} discussed the memory footprints of CPU2006 workloads, while \citet{CPU2006_WSS} analysed their working set sizes. \citet{PARSEC} present memory behavior of PARSEC benchmark suite. \citet{WorkloadCharacterizationMethods} discusses a taxonomy of workload characterization techniques.

%% file: conclusion.tex
\section{Conclusion}
\label{section:conclusion}
In this paper, we provide the first, comprehensive characterization of the 
memory behavior of the SPEC CPU2017 benchmark suite. Our working set analysis
shows that many workloads have a working set much higher than 32~MB (maximum 
cache size assumed in our experiments), implying the continued importance of 
cache hierarchies for benchmark performance. We also show that Rate benchmarks, both INT and 
FP, still have main memory consumption well below 900~MB, which was target 
memory footprint for CPU2006. Almost 90\% of the workloads have main memory 
consumption below 5~GB, with the average across the suite being 1.82~GB. However, workloads have extremely varying peak memory bandwidth usage, with some benchmarks requiring as little as 0.2 MB/s, to others utilizing upto 2.3 GB/s.

In addition, our experiments have revealed some interesting results with respect to dynamic instruction counts and distributions. The average instruction count for SPEC CPU2017 workloads is 22.19 trillion, which is an order of magnitude higher 
than the SPEC CPU2006. In addition, we find that FP benchmarks typically have 
much higher compute requirements: on average, FP workloads carry out
three times the number of arithmetic operations as compared to INT workloads.